\documentclass[]{aa} 
\usepackage{graphicx}
\usepackage{txfonts}
\usepackage[dvipsnames]{xcolor}
\usepackage[normalem]{ulem}  

\newcommand{\Iv}{\mathcal{I}_{\nu}}
\newcommand{\Ivlib}{\mathcal{I}_{\nu,\mathrm{lib}}}
\newcommand{\Ivcross}{\mathcal{I}_{\nu,\mathrm{cross}}}

\usepackage{siunitx/siunitx}
\DeclareSIUnit\msun{M_\odot}
\DeclareSIUnit\year{yr}
\DeclareSIUnit\au{AU}

\usepackage[backref, pagebackref=false, hyperindex=true, breaklinks=true, colorlinks=true, urlcolor=magenta, linkcolor=magenta, citecolor=NavyBlue, citebordercolor={0 1 0}, pagecolor=red, bookmarks=true, filecolor=blue,bookmarksopen=true, plainpages=false, pdfpagemode=UseThumbs]{hyperref}

\begin{document}

\title{The importance of the dynamical corotation torque for the migration of low-mass planets
}
\subtitle{1D analytical prescriptions verified by 2D hydrodynamical simulations}

\author{Jesse Weder \inst{\ref{unibe}}
        \and
        Clément Baruteau \inst{\ref{clement_affiliation}}
        \and
        Christoph Mordasini \inst{\ref{unibe}}
        }

\institute{
        Division of Space Research and Planetary Sciences, Physics Institute, University of Bern, Gesellschaftsstrasse 6, 3012 Bern, Switzerland\\ \email{jesse.weder@unibe.ch} \label{unibe}
        \and
        IRAP, Universit{\'e} de Toulouse, CNRS, Université Paul Sabatier, CNES, Toulouse, France \label{clement_affiliation}
        }

\date{Received March 18, 2025; accepted June 27, 2025}

\abstract
{Recent developments in the evolution of protoplanetary discs have suggested that planet formation occurs in regions of the discs with low levels of turbulent viscosity. In such environments, the dynamical corotation torque is thought to play an important role by slowing down the migration of low-mass planets (type I migration).
}
{We aim to provide a simple analytical prescription for the dynamical corotation torque for use in 1D global models of planet formation and evolution, and assess the importance of the dynamical corotation torque for the migration of low-mass planets in low-viscosity discs.
}
{We propose simple prescriptions for calculating in 1D the time evolution of the vortensities of the librating and orbit-crossing flows around a low-mass planet, which both enter the analytical expression for the dynamical corotation torque. One of our prescriptions involves a memory timescale for the librating flow, and 2D hydrodynamical simulations of disc-planet interactions are used to assess the memory timescale and validate our model.
}
{The orbital evolution of a low-mass planet is calculated by 1D simulations where the dynamical corotation torque features our prescriptions for the vortensities of the librating and orbit-crossing flows, and by 2D hydrodynamical simulations of disc-planet interactions, assuming locally isothermal discs. We find very good agreement between the 1D and 2D simulations for a wide parameter space, whether the dynamical corotation torque slows down or accelerates inward migration. We provide maps showing how much the dynamical corotation torque reduces the classical type I migration torque as a function of planet mass and orbital distance. The reduction is about 50\% for a 10 Earth-mass planet at 10 au in a young disc with surface density profile in $r^{-1/2}$ and alpha viscosity of $10^{-4}$.
}
{
In discs with low turbulent viscosity, the dynamical corotation torque should be taken into account in global models of planet formation and evolution as it can strongly slow down type I migration.
}


\maketitle
\nolinenumbers
\section{Introduction}
Understanding planet formation and evolution has remained a difficult endeavour as many physical processes act simultaneously on different length and time scales. The use of global models in planetary population syntheses helps assess the interaction between these processes (see, e.g., the review by \citealp{BurnMordasini24}), but such global models rely on a good understanding of the underlying individual processes and on having simple prescriptions that still capture their physical nature. One fundamental ingredient of global models of planet formation and evolution is the evolution of the protoplanetary disc, as it impacts how planets grow and migrate. Past models usually assumed discs with alpha turbulent viscosities $\alpha_\mathrm{vis} \gtrsim \mathrm{a\;few}\times10^{-3}$ \citep{bitsch_growth_2015,liu_super-earth_2019,Emsenhuber2021a,drazkowska_planet_2023}. However, observational constraints on turbulent line broadening from the gas emission of protoplanetary discs, or on vertical/radial diffusion of dust from the modelling of discs continuum emission, indicate that $\alpha_\mathrm{vis} \lesssim 10^{-3}$ in the outer part of discs (at a few tens of au, see review by \citealp{pinte_kinematic_2023}). These values of $\alpha_\mathrm{vis}$ are not consistent with what is needed to explain observed stellar accretion rates by angular momentum transport through turbulence \citep{Hartmann1998,Emsenhuber2023}. This and the lack of physical processes to sustain such high levels of turbulence throughout the disc are part of the reason why MHD winds have gained popularity as a way to transport angular momentum and drive disc evolution \citep{Bai2013,Bai2013a}. Although a protoplanetary disc that is threaded by a magnetic field is susceptible to the magneto-rotational instability (MRI),  detailed modelling including non-ideal MHD effects showed that MRI is suppressed in large parts of the disc from one to several tens of astronomical units (au). Turbulence in this region is thus mostly set by hydrodynamical instabilities and is expected to imply $\alpha_\mathrm{vis} \lesssim 10^{-4}$ \citep[see review by][]{Lesur2022}. This coincides with the region where planet formation is thought to take place.

With this in mind it has become very relevant to investigate planet formation and evolution in MHD wind-driven discs at low effective turbulence $\alpha_\mathrm{vis}$ \cite[e.g.,][]{Ogihara2015,Ogihara2018,Kimmig2020,kimura_predicted_2022,wafflard-fernandez_planet-disk-wind_2023,lau_sequential_2024}. However, moving towards low turbulent viscosities comes with many challenges for planet formation and evolution models as the absence of viscous heating leads to the loss of convergence zones for migration \citep{Emsenhuber2021a} and the absence of diffusion leads to a saturation of the corotation torque, ultimately leading to faster, inward-only type I migration. The presence of an MHD wind promotes flatter or even inverted surface density profiles in the inner disc which slows down or even reverses inward migration \citep{Ogihara2015}. Further moving towards low turbulent viscosities also promotes the dynamical corotation torque which can reduce \citep{paardekooper_dynamical_2014,ogihara_effects_2017,Ndugu2021} or, conversely, also speed up type I migration \citep{paardekooper_dynamical_2014,pierens_fast_2015}.

In this work we study the importance of the dynamical corotation torque on type I migration in low-viscosity discs. In Section \ref{sec:fund_analytical_theory} we propose a simple description of the dynamical corotation torque for use in 1D discs models. In Section \ref{sec:2D_simulations} we run 2D hydrodynamic simulations of disc-planet interactions to investigate the effect of vortensity mixing in the planet's horseshoe region due to viscosity, which provides parameterisations needed for our 1D model. In Section \ref{sec:1D_simulations} we verify our model by direct comparison between 2D hydrodynamical simulations and 1D simulations. This enables us to assess the importance of the dynamical corotation torque on planet formation in Section \ref{sec:influence_on_planet_formation}. A discussion follows in Section~\ref{sec:discussion}, which includes notably an alternative parametrisation of the dynamical corotation torque for 1D models, before we conclude.

\section{Dynamical corotation torque: preliminaries and memory timescale}
\label{sec:fund_analytical_theory}

Classically, planet migration is assessed by calculating the disc torque onto a planet held on a fixed orbit. In 2D non-magnetised disc models, the torque comprises\footnote{Thermal torques that arise from accretion onto the planet \citep{masset_coorbital_2017} and dust-related torques \citep{benitez-llambay_torques_2018,guilera_quantifying_2023,chrenko_pebble-driven_2024} are discarded here for simplicity.} the Lindblad torque ($\Gamma_\mathrm{lin}$, angular momentum exchange between the planet and its wakes in the disc) and the static corotation torque ($\Gamma_\mathrm{c,static}$, angular momentum exchange between the planet and the disc in its horseshoe region; hereafter HS region). Several studies, including \cite{Paardekooper2011}, have studied the static corotation torque for planets on fixed orbits. In particular, when the disc's turbulent viscosity is high enough, the static corotation torque remains unsaturated whereas it saturates at low viscosity, leaving only the Lindblad torque. 

\begin{figure}[h]
    \centering
    \includegraphics[width=1.0\linewidth]{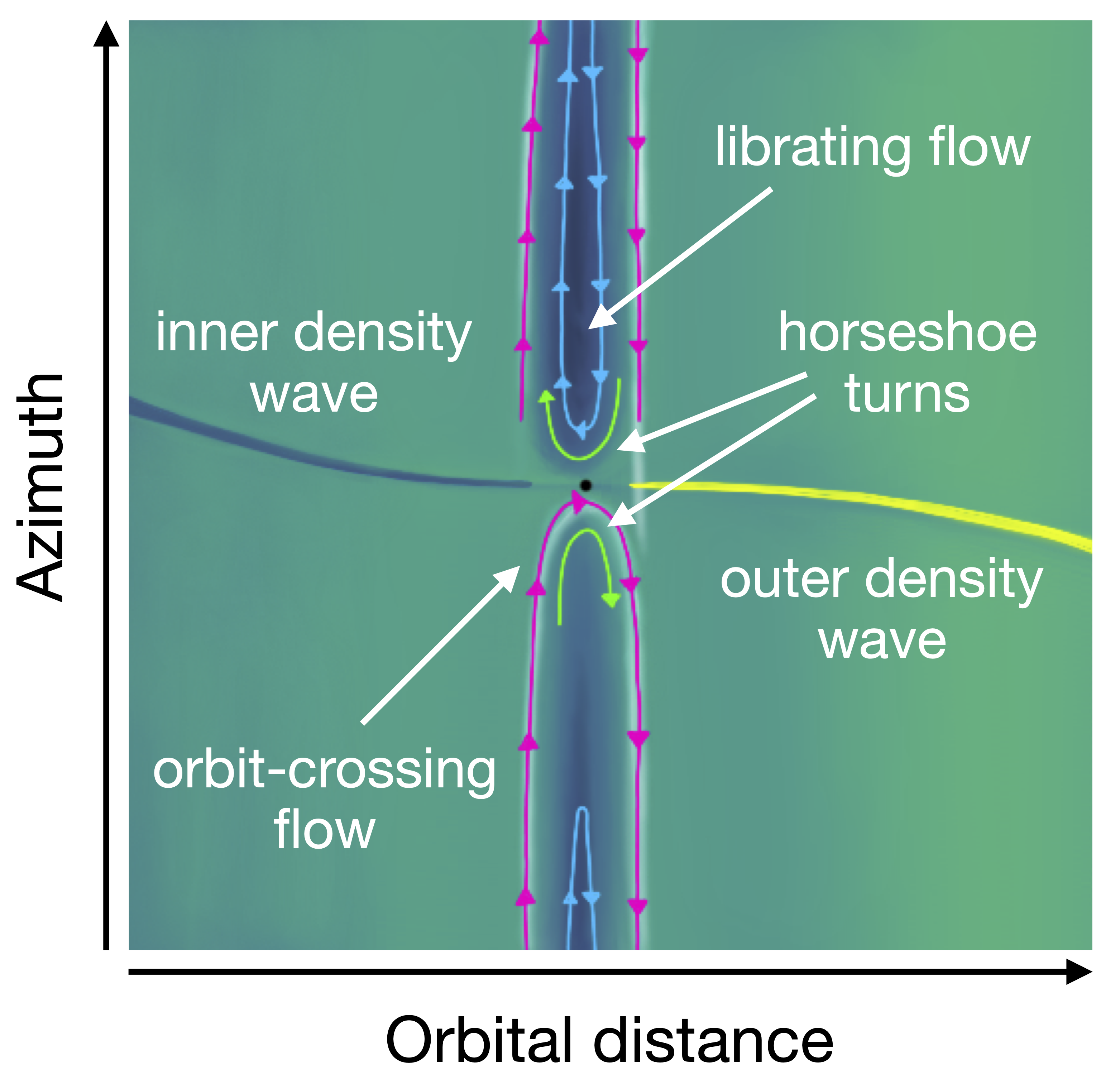}
    \caption{Sketch illustrating the perturbed vortensity around the planet's vicinity (black circle) with relevant flows and torques being highlighted: the inner and outer spiral density waves inducing the Lindblad torque, horseshoe U-turns ahead and behind the planet leading to the static corotation torque, and the librating and orbit-crossing flows giving rise to the dynamical corotation torque.}
    \label{fig:sketch}
\end{figure}

When the planet and the disc drift relative to each other, the corotation torque actually features an additional component called dynamical corotation torque ($\Gamma_\mathrm{c,dyn}$). The latter itself has two contributions \citep[see review by][]{Baruteau2016}. One is due to gas just outside the planet's horseshoe region, which executes a single U-turn with respect to the migrating planet. The torque due to this orbit-crossing flow scales with the planet's migration rate and has a positive feedback on migration. The second contribution is from the gas that remains trapped in the librating region of the planet as the latter migrates. The torque due to this librating flow also scales with the planet's migration rate, but has a negative feedback on migration. Whether the dynamical corotation torque has a net positive or negative feedback on migration amounts to comparing the (inverse) vortensity of the librating and orbit-crossing flows \citep{masset_runaway_2003}. Figure \ref{fig:sketch} provides an illustration of the various flows around the planet and the associated torques.

Following \cite{paardekooper_dynamical_2014} and \cite{mcnally_low_2017} the dynamical corotation torque can be written as
\begin{equation}\label{eq:dyn_crt_analytic}
    \Gamma_\mathrm{c,dyn} = \frac{3}{4}r_\mathrm{P}\Omega_\mathrm{P}^3\int_{y_s}^{x_\mathrm{s}} \left( \Ivlib - \Ivcross \right) x^2 dx,
\end{equation}
where $r_\mathrm{P}$ and $\Omega_\mathrm{P}$ denote the planet's orbital radius and its Keplerian frequency, $\Ivlib$ and $\Ivcross$ the inverse vortensity of the librating and orbit-crossing flows. The quantity $\Iv$ is defined as $\Iv = \Sigma / \{(\vec{\nabla} \times \vec{v}) \cdot \vec{e_z}\}$, with $\Sigma$ the surface density of the disc, $\vec{v}$ its velocity field and $\vec{e_z}$ the unit vector along the vertical direction. The relative motion between the disc and the planet implies that the planet's horseshoe region is asymmetric. This is reflected by the bounds in the above integral, which feature $x_\mathrm{s}$, the radial half-width of the horseshoe region, and the quantity $y_s = x_\mathrm{s} + \dot{r}_\mathrm{P} \tau_\mathrm{lib}/2$, with $\dot{r}_\mathrm{P}$ the planet's migration rate and $\tau_\mathrm{lib} = 8\pi r_\mathrm{P}/(3 \Omega_\mathrm{P}x_\mathrm{s})$ the libration timescale \cite[see Figure 2 in][]{paardekooper_dynamical_2014}. In the present work we neglect the radial velocity of the background gas, in contrast, for instance, to \cite{mcnally_low_2017}.

As recalled above, the dynamical corotation torque depends on the difference in vortensity between the librating and the orbit-crossing flows. In the following sections, we propose a simple descriptive model for estimating both vortensities, assuming an inwardly migrating low-mass planet. In particular, we propose a semi-analytical model for the vortensity of the librating flow.

\subsection{Inverse vortensity of the orbit-crossing flow} 
\label{sec:Ivcross}

For a planet that migrates inwards, the orbit-crossing flow originates at the radial distance $r_\mathrm{P} - x_\mathrm{s}$. For a low-mass planet such that its wakes shock the disc far enough from the planet to not influence the horseshoe region, which is the case for planet-to-star mass ratios $q_\mathrm{P} \ll h^3(r_\mathrm{P})$ with $h$ the disc's aspect ratio \citep{Lin_Papaloizou_1993}, the inverse vortensity of the orbit-crossing flow takes its local, unperturbed value. This results in
\begin{equation}\label{eq:Ivcross}
    \Ivcross = \Iv (r_\mathrm{P} - x_\mathrm{s}) \approx \Iv(r_\mathrm{P}) \times \left[1-\frac{x_\mathrm{s}}{r_\mathrm{P}} \left( \frac{3}{2} -\beta \right) \right],
\end{equation}
where we have assumed that the disc's surface density $\Sigma \propto r^{-\beta}$ and that the disc's velocity is purely Keplerian\footnote{The vortensity of a Keplerian disc is $ \frac{\{\vec{\nabla} \times \vec{v}\} \cdot e_z}{\Sigma} = \frac{1}{r\Sigma}\frac{\partial}{\partial r}(r^2\Omega) = \frac{\Omega}{2\Sigma}$.}. The half-width of the planet's horseshoe region is computed as $x_{\rm s} = Cr_{\rm P} \sqrt{q_{\rm P} / h(r_{\rm P})}$ with $C$ a constant factor that varies with the softening length of the planet's gravitational potential (see Eq. 49 in \citealp{Paardekooper2011}).

\subsection{Inverse vortensity of the librating flow}
\label{sec:memory_sect2}

The gas vortensity satisfies an advection-diffusion equation with an additional source term in $(\nabla \Sigma \times \nabla P) / \Sigma^3$ ($P$ denotes pressure) that represents baroclinic generation of vortensity \cite[see, e.g., Eq. 22 in][]{baruteau_recent_2013}. For a locally isothermal disc, the baroclinic source term is proportional to $dT/dr\times\partial\Sigma / \partial\varphi$ and vortensity can be generated across the wakes of the planet \citep{casoli_horseshoe_2009,ziampras_migration_2024}. In the absence of baroclinic source term, which is the case for barotropic flows (such as in globally isothermal discs), vortensity evolves solely through advection-diffusion. The vortensity of the librating flow then depends on the migration history of the planet and how fast turbulent viscosity mixes the librating vortensity with that of the orbit-crossing flow. We introduce a memory timescale, $\tau_\mathrm{memory}$, which is a measure of the time it takes for the librating vortensity to adapt to the vortensity of the ambient gas. A planet that migrates in a disc with non-zero vortensity gradient carries the vortensity from its past locations. We assume that at a given time $t$, the planet carries the vortensity it had at time $t - \tau_\mathrm{memory}$, when it was located at a radial distance $r_\mathrm{P}-\dot{r}_\mathrm{P} \tau_\mathrm{memory}$. More specifically, we will assume that
\begin{equation}
    \Ivlib = \Iv( \min [ r_\mathrm{P}-\dot{r}_\mathrm{P} \tau_\mathrm{memory},r_0 ] ),
    \label{eq:Ivlib_prelim}
\end{equation}
where the minimum function $\min[]$ ensures that the vortensity cannot originate from further away than the planet's initial location $r_0$. As mentioned already, in this work we limit ourselves to inward migrating planets ($\dot{r}_\mathrm{P}<0$), and for outward migration ($\dot{r}_\mathrm{P}>0$) one should replace the minimum function in Eq.~(\ref{eq:Ivlib_prelim}) by the maximum one. Introducing the migration timescale $\tau_\mathrm{mig}$ such that $\dot{r}_\mathrm{P} = r_\mathrm{P}/\tau_\mathrm{mig}$, one gets
\begin{eqnarray}
	\Ivlib &=& \Iv( \min[ r_\mathrm{P} \times (1 - \tau_\mathrm{memory}/\tau_\mathrm{mig}) ,r_0] ) \\
        &=& \Iv( r_\mathrm{P} \times \min \left[ 1-\frac{\tau_\mathrm{memory}}{\tau_\mathrm{mig}}, \frac{r_0}{r_\mathrm{P}} \right] ) \\
        &=& \Iv(r_\mathrm{P}) \times \left( \min\left[ 1-\frac{\tau_\mathrm{memory}}{\tau_\mathrm{mig}}, \frac{r_0}{r_\mathrm{P}} \right] \right)^{3/2 - \beta}. \label{eq:Ivlib}
\end{eqnarray}
Viscous diffusion will take $\Ivlib$ towards the local unperturbed vortensity $\Iv(r_\mathrm{P})$ on a timescale similar to the viscous timescale over the HS region $\tau_\mathrm{vis}=x_\mathrm{s}^2/\nu$. For high turbulent viscosity $\nu$, the viscous timescale is short compared to the migration timescale and $\Ivlib$ is close to local unperturbed value $\Iv(r_\mathrm{P})$, making the dynamical corotation torque ineffective. For low turbulent viscosity, the viscous timescale across the HS region is long compared to the migration timescale and $\Ivlib$ deviates progressively from the local unperturbed value so that the dynamical corotation torque is expected to increase in magnitude. Note that in the limiting case of vanishing turbulent viscosity, the planet carries along the vortensity at its initial location. Further, it can be seen that for a disc with zero vortensity gradient ($\beta=3/2$) the dynamical corotation torque vanishes. For a planet migrating up/down a vortensity gradient (migrating into regions with higher/lower vortensity), the dynamical corotation torque will slow down/speed up migration. The reader is referred to \cite{paardekooper_dynamical_2014} for an in-depth discussion of the different regimes of the dynamical corotation torque.

The mixing process of vortensity in the planet's HS region is complex and still needs to be understood in detail. In our model, this process is modelled by a single parameter $\tau_\mathrm{memory}$. We expect $\tau_\mathrm{vis}$ to pose an upper limit on $\tau_\mathrm{memory}$. However, this needs to be confirmed by measuring the vortensity evolution in 2D hydrodynamical simulations. In the following, we present results of 2D hydrodynamical simulations in order to get insight into the vortensity evolution around migrating planets.

\section{2D hydrodynamical simulations}
\label{sec:2D_simulations}

\subsection{Numerical setup}
\label{sec:setup_2D_simulations}

We run 2D hydrodynamical simulations of disc-planet interactions. This involves solving the continuity and Navier-Stokes equations, which are closed by an ideal, locally isothermal equation of state \cite[see][for the detailed equations]{masset_co-orbital_2002}. We use \texttt{FARGO3D} \citep{benitez-llambay_fargo3d_2016} along with the orbital advection algorithm of \cite{masset_fargo_2000} to solve these equations. Calculations are performed on a cylindrical grid $[r,\phi]$ with $r\in [0.4,2.5]r_0$ and $\phi \in[-\pi,\pi]$, where $r_0$, the planet's initial orbital radius, defines the code's unit of length. We use $N_r=782$ cells linearly or logarithmically spaced along $r$, and $N_\phi=2346$ cells linearly spaced along $\phi$. This choice of $N_r$ allows to resolve the horseshoe region at the planet's initial location by 10 cells. For the boundary conditions at the grid's radial edges, we apply a power-law extrapolation for the surface density, a Keplerian extrapolation for the azimuthal velocity, and an antisymmetric boundary condition for the radial velocity (thereby ensuring that the radial velocity cancels out at the grid's radial edges). We also set wave-damping zones for $r / r_0 \in [0.4,0.52]\cup[1.92,2.5]$ to avoid reflections of the planet wakes \cite[see][]{de_val-borro_comparative_2006}.

The gas surface density is initialised as a power-law, $\Sigma_0(r) = \Sigma_0 (r/r_0)^{-\beta}$, with $\Sigma_0$ the initial surface density at the planet's initial location $r_0$. We assume the disc evolution is driven by radial transport of angular momentum and is parameterised by the classical $\alpha$-viscosity approach \citep[][the alpha viscosity is taken uniform and denoted by $\alpha_\mathrm{vis}$]{Shakura1973}. The disc's aspect ratio is given by $h(r)=h_0(r/r_0)^f$ with $h_0 = h(r_0)$. In our fiducial simulations we adopt $h_0=0.05$ and $f=0.5$. This implies a uniform background temperature, which avoids sharp vortensity perturbations close to the planet that would be produced by the baroclinic source term $\propto\partial_{\phi}\Sigma\,\partial_rT$ in the vortensity equation (see Section \ref{sec:memory_sect2}). Results of simulations with a non-uniform temperature profile are presented in Section~\ref{sec:disc_2Dsimulations} and additional globally isothermal simulations with $h_0=0.07$ are shown in Appendix \ref{app:high_aspect_ratio}.

A freely moving planet is introduced at the beginning of our simulations. The planet-to-star mass ratio $q_\mathrm{P}=M_\mathrm{P}/M_\star=10^{-5}$, which corresponds to $M_\mathrm{P} \simeq 3.3\mathrm{M}_\oplus$ ($\mathrm{M}_\oplus$ corresponds to the mass of the Earth) around a solar mass star. This is about the mass of a super Earth, which is not expected to open a gap in our disc model as $q_\mathrm{P}/h^3 \ll 1$. We use a smoothing length of $0.6 H(r_\mathrm{P})$ for the gravitational potential of the planet \citep{muller_treating_2012}. To avoid spurious inward acceleration of the inwardly migrating planet due to the fact that self-gravity is discarded, the force exerted by each grid cell on the planet involves the disc's local surface density subtracted by its azimuthal average \citep{baruteau_type_2008, benitez-llambay_fargo3d_2016}. This was done in our \texttt{FARGO3D} runs by activating the \texttt{BM08} option in our code's setup. 

Since the star stays at the origin of the reference frame, indirect terms are included. These are the indirect term of the planet exerted on the disc, and the indirect term of the disc exerted on the planet. Since the disc is not self-gravitating, the indirect term of the disc on itself is not included \citep{crida_use_2022}.

Our simulations aim to investigate how planet migration impacts the evolution of the vortensity of the librating and orbit-crossing flows, for different values of $\alpha_\mathrm{vis}$, $\beta$ and $\Sigma_0$. The latter is varied according to the initial Toomre Q-parameter at the planet's initial location:
\begin{equation}
    Q_0 = \frac{h_0 r_0 \Omega^2(r_0)}{\pi G \Sigma_0}.
    \end{equation}
Our default value for $Q_0$ is 8, for which we ran simulations with $\alpha_\mathrm{vis}\in\{10^{-3},10^{-4},10^{-5}\}$ and $\beta \in \{\frac{1}{2},1,\frac{3}{2},2\}$. Note that in a locally isothermal disc, $\beta < 3/2$ implies that the dynamical corotation torque has a negative feedback on migration and therefore slows down inward migration, whereas for $\beta > 3/2$ the dynamical corotation torque has a positive feedback on migration and accelerates inward migration \citep{paardekooper_dynamical_2014}. Furthermore, we ran additional simulations with $Q_0\in \{12,16,24,32\}$, $\alpha_\mathrm{vis}\in\{10^{-4},10^{-5}\}$ and $\beta \in \{\frac{1}{2},1\}$ to get more insight into the negative feedback effect of the dynamical corotation torque. Table~\ref{tab:fargo_simlist} shows the full set of simulations presented in Section~\ref{sec:results}. Simulations were run for 3000 orbits or were stopped when the planet reached $\mathrm{r_p} \lesssim 0.5 r_0$. 

In the following, results of simulations will be expressed in code's units (c.u.), where the code's unit of mass is the mass of the star, the code's unit of length is the initial orbital radius ($r_0$) of the planet, as already stated above, and the code's unit of time is $\Omega^{-1}(r_0)$.

\begin{table}
    \begin{center}
    \caption{Summary of initial parameters explored in the 2D hydrodynamic simulations presented in Section~\ref{sec:results} ($Q_0$: Toomre-Q parameter at the planet's initial orbital radius, $\alpha_\mathrm{vis}$: turbulent alpha viscosity, $\beta = -d\log\Sigma_0 / d\log r$).}
    \label{tab:fargo_simlist}
        \begin{tabular}{c|r|c|l}
            \hline\hline
            $Q_0$ & $\alpha_\mathrm{vis}$ & $\beta$ & grid's radial spacing \\ 
            \hline
             8 & $\{ 10^{-3},10^{-4},10^{-5} \}$ & $\{\frac{1}{2},1\}$ & linear \\
               & $\{ 10^{-3},10^{-4},10^{-5} \}$ & $\{\frac{3}{2},2\}$ & logarithmic \\
            12 & $\{ 10^{-4},10^{-5} \}$ & $\{\frac{1}{2},1\}$ & linear \\
            16 & $\{ 10^{-4},10^{-5} \}$ & $\{\frac{1}{2},1\}$ & linear \\
            24 & $\{ 10^{-4},10^{-5} \}$ & $\{\frac{1}{2},1\}$ & linear \\
            32 & $\{ 10^{-4},10^{-5} \}$ & $\{\frac{1}{2},1\}$ & linear \\
            \hline
        \end{tabular}
    \end{center}
\end{table}

\subsection{Results}
\label{sec:results}
In the following sections we discuss the results of the 2D simulations (left column of Figure \ref{fig:simulations_TQ8_migrate_orbital_evo}). The 1D simulations are introduced and discussed later in Section \ref{sec:1D_simulations}.

\subsubsection{Migration rates and orbital distances}
\label{sec:fargo_mig_rates_orbit_evo}

\begin{figure*}[]
    \centering
    \includegraphics[trim={1cm 1.5cm 2cm 0},clip,width=\hsize]{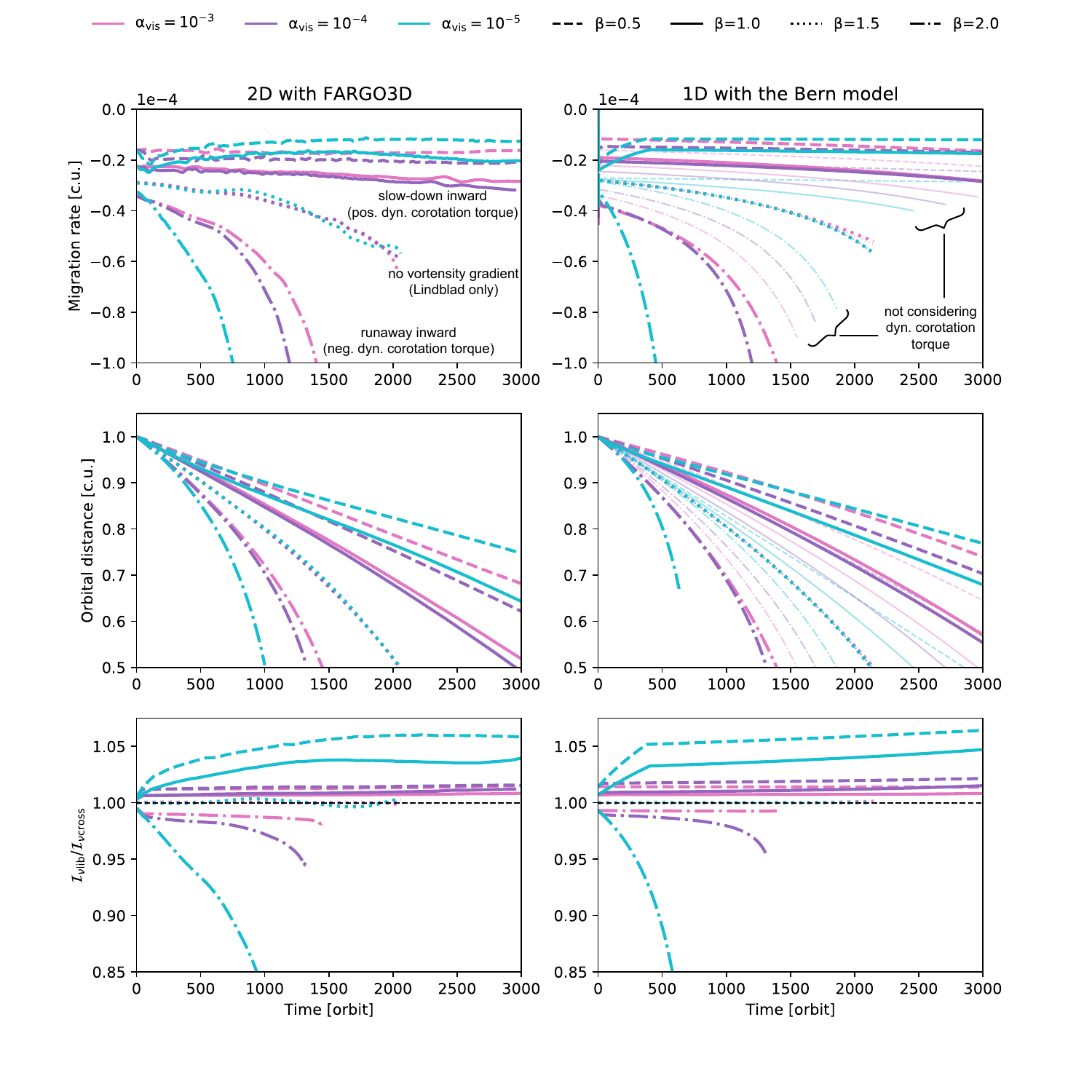}
    \caption{From top to bottom: time evolution of the planet's migration rate, orbital distance, and inverse vortensity ratio $\Ivlib / \Ivcross$ for simulations with $Q_0=8$. The results of our 2D (with \texttt{FARGO3D}) and 1D (with the Bern model) simulations are shown side by side for easy direct comparison. Lines are coloured by $\alpha_\mathrm{vis}$ and different linestyles indicate varying slopes of the surface density, $\beta$. The migration rates of the 2D simulations (top-left panel) are shown as moving averages over 100 orbits. For the 1D simulations (right), we additionally show as transparent lines results of simulations that discard the dynamical corotation torque. For an easier direct comparison of the most relevant simulations, see left column in Figure \ref{fig:direct_comparison} in Appendix \ref{app:direct_compar}.}
    \label{fig:simulations_TQ8_migrate_orbital_evo}
\end{figure*}

First we present the results of simulations with $Q_0=8$ for various values of $\alpha_\mathrm{vis}$ and $\beta$ in order to investigate the different migration behaviours predicted by theory (see Section~\ref{sec:fund_analytical_theory}). In the top two panels of the left column of Figure \ref{fig:simulations_TQ8_migrate_orbital_evo} we show the time evolution of the planet's migration rate and orbital distance. For a disc with no vortensity gradient ($\beta=3/2$) the migration rate slowly increases as the planet moves inward, which simply reflects the radial dependence of the Lindblad torque since both the static and dynamical corotation torques vanish. The planet reaches the inner edge of the grid in $\gtrsim 2000$ orbits.

For negative vortensity gradients ($\beta=1/2$ and $\beta=1$), migration rates take smaller values than for $\beta=3/2$ since the static and dynamical corotation torques are both positive. We note that decreasing $\alpha_\mathrm{vis}$ from $10^{-3}$ to $10^{-4}$ leads to slightly higher migration rates, and then going further down to $\alpha_\mathrm{vis}=10^{-5}$ reduces migrations rate more significantly. This behaviour reflects how the static and dynamical parts of the corotation torque vary with viscosity, since decreasing viscosity decreases the static part but increases the dynamical part. Recall that for the range of (low) viscosities probed in our simulations, the Lindblad torque is independent of viscosity \citep{Muto09}. Only at low $\alpha_\mathrm{vis}$ and low $\beta$ does the migration rate increase with time, more appreciably at the beginning of the simulations, highlighting the negative feedback of the dynamical corotation torque on migration.

Finally, for a positive vortensity gradient ($\beta=2$), we observe runaway migration at a rate that increases with decreasing viscosity. This behaviour cannot be attributed to the static corotation torque, although it is negative here, since its magnitude decreases with decreasing viscosity. It is therefore due to the dynamical corotation torque. In these simulations, the planet reaches the grid's inner edge in 1000 to 1500 orbits. This all reflects the theoretical expectations.

\subsubsection{Vortensity of the orbit-crossing and librating flows}
\label{sec:vortensity_ratio}
We are interested in the time evolution of $\Ivlib - \Ivcross$ as it is what sets the dynamical corotation torque (see Eq.~\ref{eq:dyn_crt_analytic}). To measure $\Ivlib$ and $\Ivcross$ in our 2D simulations, we use a very similar approach to that of \cite{wafflard-fernandez_intermittent_2020}. For the orbit-crossing flow, since in our models the horseshoe U-turn timescale is short in comparison to the viscous diffusion timescale over the HS region, $\Ivcross$ can be simply measured at the orbital radius of the planet and just behind it in azimuth (as the planet migrates inwards). In practice, it is averaged over $\phi -\phi_{\mathrm p} \in [-0.2,-0.1]$ at $r = r_{\mathrm p}$ ($\phi_{\mathrm p}$ denotes the planet azimuth, which is zero in our simulations since our runs are carried out in the frame corotating with the planet). Now for the librating flow, two strategies are employed. When $\beta < 3/2$ (no runaway), $\Ivlib$ is averaged over azimuth at the planet's orbital radius, excluding a small region around the planet: in practice, it is averaged over $\phi -\phi_{\mathrm p} \in [-\pi,-0.4]\cup[0.4,\pi]$ at $r = r_{\mathrm p}$. Finally, when the planet goes into runaway migration ($\beta=2$), the librating region gets heavily distorted, leaving just an island of librating material in front of the planet in the azimuthal direction (as the planet moves inward). In that case, again in practice $\Ivlib$ is averaged over $\phi -\phi_{\mathrm p} \in [0.4,0.6]$ at $r = r_{\mathrm p}$. 

\begin{figure*}
    \centering
    \includegraphics[width=\hsize]{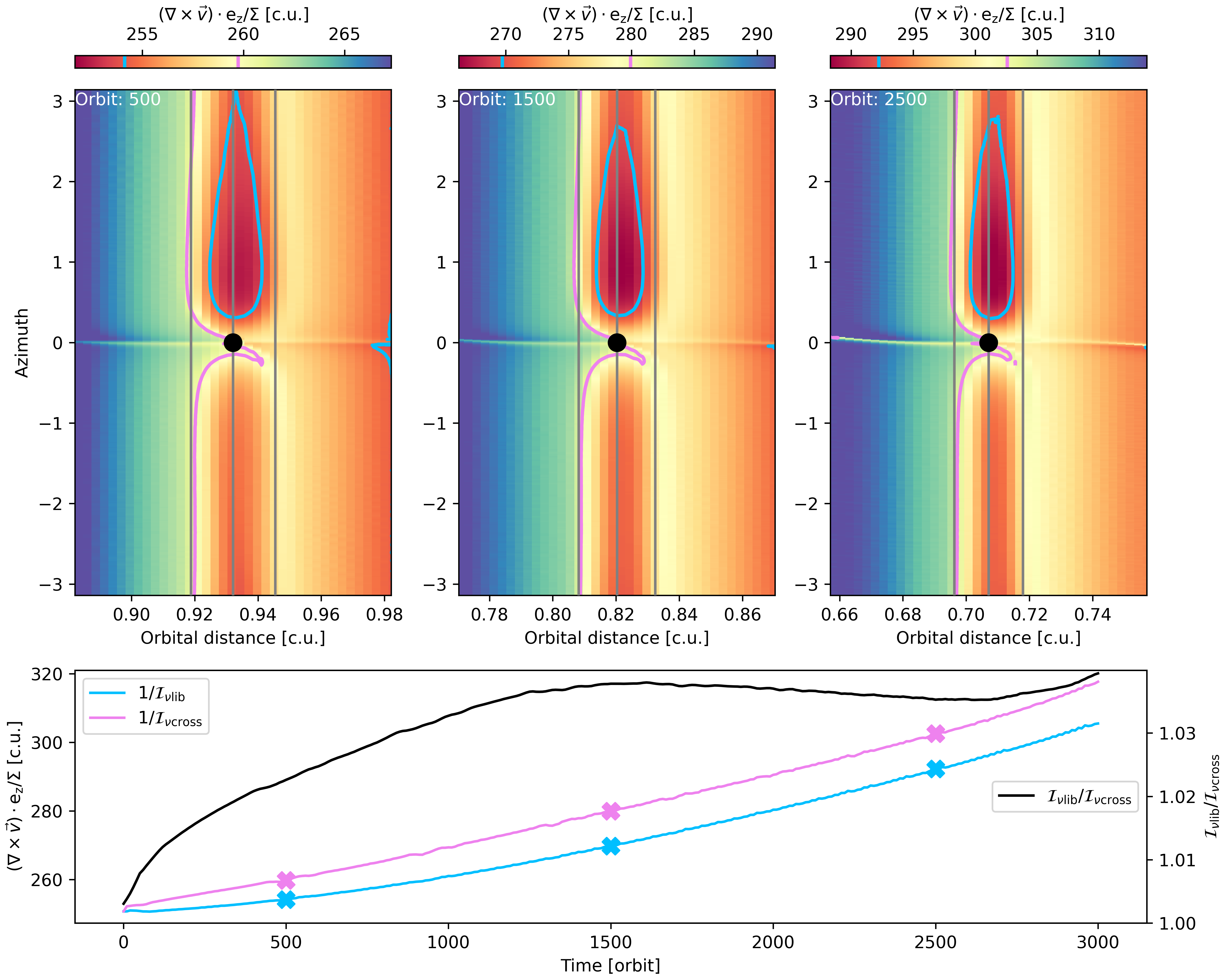}
    \caption{Evolution of the disc vortensity for the simulation with $Q_0=8$, $\beta=1$ and $\alpha_\mathrm{vis}=10^{-5}$. The top panels show snapshots of the vortensity field around the planet at 500, 1500 and 2500 orbits. The vortensity deficit in the HS region, and the vortensity advection of the orbit-crossing flow from the inner to the outer edges of the HS region are clearly visible. The vortensities of the orbit-crossing and librating flows measured by the method described in Section~\ref{sec:vortensity_ratio} are shown as contour lines in pink and cyan, respectively. The vertical grey lines mark the orbital radius of the planet and the extent of its horseshoe region (see Section~\ref{sec:Ivcross} for the expression of the horseshoe half-width $x_{\rm s}$). The bottom panel shows the time evolution of the measured vortensities for the orbit-crossing and librating flows. Crosses mark the results shown in the top panels. The time evolution of $\Ivlib / \Ivcross$ is shown in black.}
    \label{fig:fargo_2Dplot}
\end{figure*}

The upper panels in Figure \ref{fig:fargo_2Dplot} illustrate the above method for a simulation with $\beta=1$, in which case we see that the evolution of the vortensities is well captured throughout the simulation. Note the orbit crossing flow originating from $\simeq r_\mathrm{P}-x_\mathrm{s}$ which is in agreement with Eq. \ref{eq:Ivcross}. In the lower panel of the figure, we stress that, although both vortensities increase with time, the ratio $\Ivlib / \Ivcross$ reaches a quasi-steady state after $\sim 1500$ orbits. The ratio is not necessarily expected to remain at a constant value as the migration rate and the width of the HS region vary with orbital distance.

Back to Figure \ref{fig:simulations_TQ8_migrate_orbital_evo}, the lower-left panel shows the time evolution of $\Ivlib / \Ivcross$ in the simulations. For $\beta < 3/2$, the planet moves up the disc's vortensity gradient and thus has a vortensity deficit in its HS region, which explains why $\Ivlib / \Ivcross > 1$. Quite remarkably, in all our simulations with $\beta < 3/2$, $\Ivlib / \Ivcross$ reaches a near-stationary value while the planets keeps migrating, after a characteristic timescale that depends on $\beta$ and $\alpha_\mathrm{vis}$. The lower the viscosity, the longer it takes to reach a steady state, the higher the final value for $\Ivlib / \Ivcross$ and the stronger the negative feedback of the dynamical corotation torque on migration, as already shown above. For the disc and planet parameters of these simulations, the dynamical corotation torque really starts impacting migration for $\alpha_\mathrm{vis} \lesssim 10^{-4}$. For $\beta > 3/2$, the planet moves down the vortensity gradient and thus builds up a vortensity excess in its HS region, implying that $\Ivlib / \Ivcross < 1$. This time, the runaway implies that $\Ivlib / \Ivcross$ keeps decreasing with time and does not reach a steady state value. Finally, for $\beta=3/2$ we see that $\Ivlib / \Ivcross$ stays at unity regardless of turbulent viscosity, as expected.

\subsection{Estimating the memory timescale}
\label{subsec:memory_timescale}
As shown in Section \ref{sec:vortensity_ratio}, $\Ivlib / \Ivcross$ eventually reaches a quasi-steady state when migration does not run away ($\beta < 3/2$). While for simulations with $\alpha_\mathrm{vis} \gtrsim 10^{-4}$ the final value of $\Ivlib / \Ivcross$ can be essentially attributed to the background vortensity gradient, for simulations at lower viscosity 
($\alpha_\mathrm{vis}=10^{-5}$) $\Ivlib / \Ivcross$ takes much larger values, which points to a memory effect for the vortensity of the librating flow (see Section~\ref{sec:memory_sect2}).

To estimate the memory timescale in our 2D simulations, we simply make use of Eq.~(\ref{eq:Ivlib}), assuming that $1-\tau_\mathrm{memory}/\tau_\mathrm{mig} < r_0/r_\mathrm{P}$ for $\Ivlib / \Ivcross$ to reach a constant value. Reversing Eq.~(\ref{eq:Ivlib}) then gives 
\begin{equation}\label{eq:t_memory}
    \tau_\mathrm{memory} =  \tau_\mathrm{mig} \times \left( 1 - \left( \frac{\Ivlib}{\Iv(r_\mathrm{P})} \right)^\frac{1}{3/2 - \beta}\right),
\end{equation}
where we recall that the migration timescale $\tau_\mathrm{mig}=r_\mathrm{P}/\dot{r}_\mathrm{P}$. Although migration rates are close to stationary in simulations with $\beta < 3/2$ (see Fig. \ref{fig:simulations_TQ8_migrate_orbital_evo}), the migration timescale varies over time due to the planet moving inwards, and so does the memory timescale.

\begin{figure*}[]
    \centering
    \includegraphics[trim={1cm 0 1.5cm 0},clip,width=\hsize]{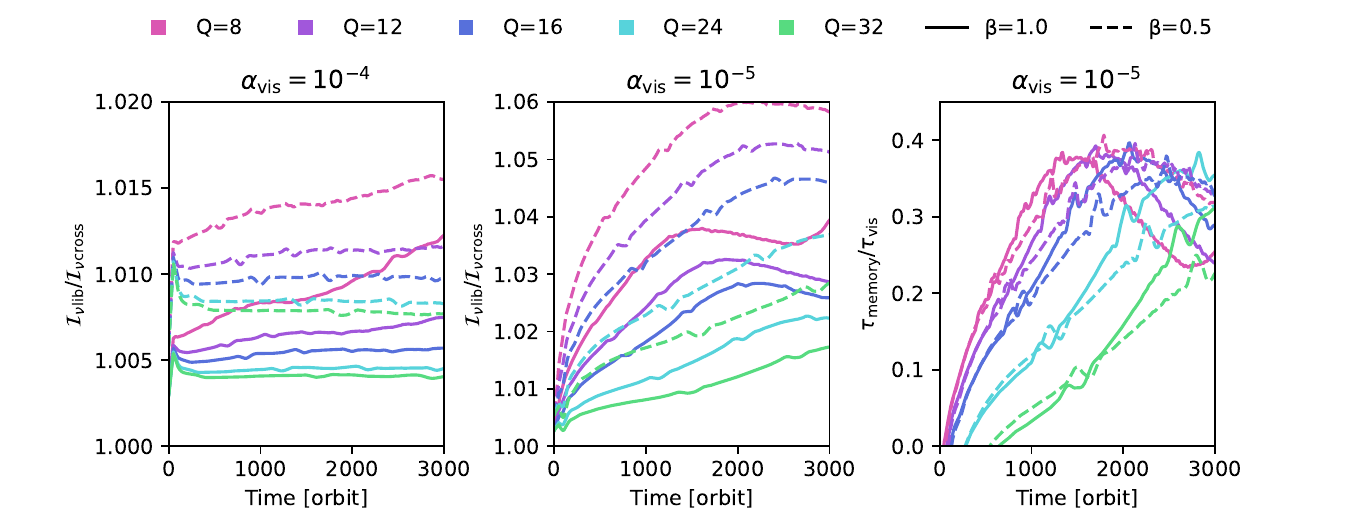}
    \caption{Left and middle panels: time evolution of $\Ivlib / \Ivcross$ in simulations with varying $Q_0\in \{8,12,16,24,32\}$, $\beta \in \{ \frac{1}{2}, 1\}$, for $\alpha_\mathrm{vis} = 10^{-4}$ (left) and $\alpha_\mathrm{vis} = 10^{-5}$ (middle). Right panel: time evolution of the ratio between the memory timescale and the viscous timescale across the HS region for the $\alpha_\mathrm{vis} = 10^{-5}$ simulations.}
    \label{fig:tau_memory}
\end{figure*}

We computed the memory timescale in an extended set of 2D simulations that include higher values for $Q_0$ (12, 16, 24 and 32), for $\beta \in \{ \frac{1}{2}, 1\}$ and $\alpha_\mathrm{vis}\in\{10^{-4},10^{-5}\}$. Results are shown in Figure~\ref{fig:tau_memory}. The left panel first shows the time evolution of $\Ivlib / \Ivcross$ for $\alpha_\mathrm{vis}=10^{-4}$. Two trends emerge: (i) for a given $Q_0$, $\Ivlib / \Ivcross$ is larger for smaller $\beta$, and (ii) for a given $\beta$, $\Ivlib / \Ivcross$ decreases with increasing $Q_0$. The viscous and libration timescales over the HS region happen to have similar values within about 30\% ($\tau_\mathrm{vis} \simeq 130$ orbital periods at $r_0$, $\tau_\mathrm{lib}\simeq 100$ orbital periods at $r_0$). This implies that turbulent viscosity takes the vortensity of the librating flow towards the local unperturbed value on a timescale similar to the libration timescale and thus no clear memory effect is expected to occur. We thus suspect that the ratio $\Ivlib / \Ivcross$ is essentially explained by the background vortensity gradient (see Eq.~\ref{eq:Ivcross} with $\Ivlib \sim \Iv(r_{\mathrm p})$). Evaluating Eq.~(\ref{eq:Ivcross}) at the planet's initial location actually gives $\Iv(r_{\mathrm p}) / \Ivcross \approx 1.014$ for $\beta=1/2$ and $\approx 1.007$ for $\beta=1$, which is in decent agreement (albeit slightly above) the values obtained in the left panel of Figure~\ref{fig:tau_memory}.

The middle panel of Figure~\ref{fig:tau_memory} now shows $\Ivlib / \Ivcross$ for $\alpha_\mathrm{vis}=10^{-5}$. Here the viscous timescale is one order of magnitude larger than the libration timescale and we observe a clear build-up of vortensity memory that reaches a quasi-steady state for sufficiently fast migration ($Q\lesssim 24$). Having previously shown that the background vortensity gradient also contributes to the final value for $\Ivlib / \Ivcross$, we have subtracted the contribution of the background vortensity gradient when calculating the memory timescale using Eq.~(\ref{eq:t_memory}). The right panel of Figure~\ref{fig:tau_memory} shows the obtained memory timescale divided by the local viscous timescale over the HS region (half-width). Note that, although simulations with different $\beta$ values reach a quasi-steady state for different values of $\Ivlib / \Ivcross$, they show very similar memory timescales. The panel also shows that the memory timescale has a mild dependence on the migration rate (via $Q_0$), but overall $\tau_\mathrm{memory} \sim [0.3-0.4] \times \tau_\mathrm{vis}$ is a good match to our simulations results. These experiments allow us to conclude that two effects impact the ratio $\Ivlib / \Ivcross$: the background vortensity gradient and a memory effect at low viscosity for the vortensity of the librating flow, which acts on a typical timescale $\simeq 0.4 \tau_\mathrm{vis}$.

\section{1D simulations}
\label{sec:1D_simulations}

In this section, we present the results of 1D simulations that model the disc evolution and the planet's orbital evolution. We compare them with the results of the 2D simulations of the previous section in order to test our new formulation for the dynamical corotation torque based on a memory timescale argument. The 1D simulations use parts of the Bern model of planet formation and evolution \citep{Alibert2005,Mordasini2012,Emsenhuber2021a}.

\subsection{Numerical setup}
\label{sec:setup_1D}

The 1D simulations model the evolution of a protoplanetary disc by solving the classical viscous diffusion equation for the disc's surface density
\begin{equation}\label{eq:disc_evo_visc}
    \frac{\partial \Sigma}{\partial t} = \frac{1}{r}\frac{\partial}{\partial r}\left[ \frac{3}{r\Omega}\frac{\partial}{\partial r}(r^2 \Sigma \alpha_\mathrm{vis} c_\mathrm{s}^2) \right],
\end{equation} 
with $c_\mathrm{s}=h r \Omega$ the sound speed and $\alpha_\mathrm{vis}$ the alpha turbulent viscosity. Like in our 2D simulations, the disc's aspect ratio $h=h_0 (r/r_0)^f$ with $f=1/2$ (uniform temperature profile), the initial surface density is $\Sigma_0 (r/r_0)^{-\beta}$, and  $\alpha_\mathrm{vis}$ is taken constant. The 1D grid uses $1244$ cells logarithmically spaced over $r\in [0.2,2.5]r_0$. For the boundary conditions, we use again a power-law extrapolation for the surface density. The default timestep is one orbit at the planet's initial location with additional constraints to ensure that the Courant-Friedrichs-Lewy condition is fulfilled and the planet does not move more than $10^{-4}r_\mathrm{P}$ per timestep. 

At the beginning of the simulations, a planet is inserted at $r_0$ with the same planet-to-star mass ratio as in our 2D simulations. The planet's semi-major axis is evolved according to the disc torques. For the Lindblad torque and the static corotation torque, we use the torque formulae of \cite{Paardekooper2011}\footnote{Recall that the torque formulae in \cite{Paardekooper2011} are for a smoothing length-to-scaleheight ratio $b/h=0.4$, while our 2D simulations employ $b/h=0.6$. The dependence of the individual torques on $b/h$ can be found in \cite{Paardekooper2010}.}. For the dynamical corotation torque, Eq.~(\ref{eq:dyn_crt_analytic}) gives
\begin{equation}
    \Gamma_\mathrm{c,dyn} = \frac{1}{4}r_\mathrm{P}\Omega_\mathrm{P}^3 (\Ivlib - \Ivcross)(x_\mathrm{s}^3 - y_s^3),
\end{equation}
assuming $\Ivlib$ and $\Ivcross$ are uniform over the HS region. We further set $y_s$ such that 
\begin{equation}
    y_s=\max(x_\mathrm{s}+\dot{r}_\mathrm{P} \tau_\mathrm{lib}/2,0),
\end{equation}
which ensures that in the fast migrating regime (runaway), the dynamical torque reaches
\begin{equation}
    \Gamma_\mathrm{c,dyn} = \frac{1}{4}r_\mathrm{P}\Omega_\mathrm{P}^3 (\Ivlib - \Ivcross) x_\mathrm{s}^3
\end{equation}
as discussed in \citet[][see their section 6.2]{mcnally_low-mass_2018}. This makes the calculation of the dynamical corotation torque valid in both the slow and fast migration regimes, thus independently of $\beta$. Finally, $\Ivcross$ is calculated using Eq.~(\ref{eq:Ivcross}) and $\Ivlib$ via Eq.~(\ref{eq:Ivlib}) with
\begin{equation}
    \tau_\mathrm{memory} = 0.4  \tau_\mathrm{vis} \times \mathrm{max}\left(1-\frac{\tau_\mathrm{lib}}{2\tau_\mathrm{vis}},0\right).
    \label{eq:final_memory_timescale}
\end{equation}
Eq.~(\ref{eq:final_memory_timescale}) ensures that a memory effect sets in only if the viscous timescale is sufficiently long compared to the libration timescale. If viscosity is indeed able to equilibrate the vortensity of gas that goes from the front to the back of the planet in the HS region, which takes about half a libration timescale, there will be no memory effect. A memory effect sets in at $\tau_\mathrm{vis} \simeq 2\tau_\mathrm{lib}$, corresponding to $\alpha_\mathrm{vis}\simeq 10^{-4}$ for our fiducial values ($q=10^{-5}$ and $h=0.05$). This coincides with the point where the static corotation torque saturates (see Fig. 2 in \citealp{Paardekooper2011} and Fig. 11 in \citealp{baruteau_recent_2013}). The maximum function in the above equation ensures that $\tau_\mathrm{memory}$ cannot be negative if the viscous timescale is shorter than half the librating timescale.

\subsection{Results}
Our results of 1D calculations are displayed in the right column of panels in Figure~\ref{fig:simulations_TQ8_migrate_orbital_evo} for $Q_0 = 8$ and the same range of values for $\beta$ (1/2,1) and $\alpha_\mathrm{vis}$ ($10^{-5},10^{-4},10^{-3}$) as in the 2D simulations. Overall, our results of 1D calculations are in very good agreement with those of the 2D simulations. In the calculations with no vortensity gradient ($\beta=3/2$), the disc torque is entirely given by the Lindblad torque and does not depend on $\alpha_\mathrm{vis}$. We note that the 2D torque starts to slightly exceed the 1D torque over long timescales ($\gtrsim$ 1500 orbits), which is probably due to the gradual onset of non-linearities in the 2D simulations brought about by the planet's wakes.

In discs with a negative vortensity gradient ($\beta<3/2$), it is interesting to notice the initial stage where the migration rate decreases in the simulations with low viscosity ($\alpha_\mathrm{vis}=10^{-5}$). It arises from the gradual increase in the dynamical corotation torque as a vortensity deficit builds up in the planet's librating region. This stage is over once the planet has migrated far enough that the memory of the initial vortensity is lost ($r_\mathrm{P} - \dot{r}_\mathrm{P} \tau_\mathrm{memory} < r_0$). A similar behaviour can actually be seen for the 2D simulations, which we can now attribute to the gradual increase in the $\Ivlib/\Ivcross$ ratio. For higher turbulent viscosities ($\alpha_\mathrm{vis}\gtrsim10^{-4}$), $\Ivlib/\Ivcross$ is almost entirely set by the background vortensity gradient as seen in Section~\ref{subsec:memory_timescale}. However, Eq.~(\ref{eq:Ivcross}) tends to somewhat overestimate $\Ivlib/\Ivcross$ compared to the 2D simulations (compare the bottom panels), which leads to slightly lower migration rates, and therefore slightly slower migration in the 1D calculations (compare the top and middle panels). Note that we are able to reproduce with the 1D model the higher migration rates obtained for $\alpha_\mathrm{vis}=10^{-4}$ compared to $\alpha_\mathrm{vis}=10^{-3}$, discussed in Section \ref{sec:fargo_mig_rates_orbit_evo}. This indicates that the analytical timescale picture indeed captures the governing physical processes.

Now turning our attention to discs with a positive vortensity gradient ($\beta>3/2$), we see that good agreement is obtained between the 1D and 2D simulations for $\alpha_\mathrm{vis}\geq 10^{-4}$. For our lowest viscosity ($\alpha_\mathrm{vis}=10^{-5}$), our 1D model tends to underestimate $\Ivlib/\Ivcross$, thereby predicting slightly faster runaway migration than in our 2D simulations. Yet, the agreement remains satisfactory and shows that we are able to model runaway type I migration driven by the dynamical corotation torque.

Before concluding this section, we point out that additional 1D calculations were carried out by discarding the dynamical corotation torque. The results of these 1D calculations, in which the disc torque only comprises the Lindblad and static corotation torques, are shown as transparent curves in the right panels of Figure~\ref{fig:simulations_TQ8_migrate_orbital_evo}. These extra simulations highlight the importance of including the dynamical corotation torque for the disc models we have considered. More generally, our results show that the dynamical corotation torque must be taken into account in models of low-mass planet formation, which we further emphasise in the next section.

\section{How much does the dynamical corotation torque reduce the type I migration torque?}
\label{sec:influence_on_planet_formation}
The two previous sections demonstrate that our new formulation for the dynamical corotation torque, which is based on a memory timescale for the evolution of the librating flow's vortensity in the planet's horseshoe region, reproduces well the migration behaviour of a low-mass planet obtained in 2D hydrodynamical simulations. Eqs.~(\ref{eq:Ivlib}) and~(\ref{eq:final_memory_timescale}) show that the amplitude of the dynamical corotation torque depends on how the viscous timescale across the HS region ($\tau_\mathrm{vis}$) compares with the migration timescale ($\tau_{\rm mig}$) and the libration timescale ($ \tau_{\rm lib}$), through the ratios $\tau_\mathrm{vis} / \tau_{\rm mig}$ and $\tau_\mathrm{vis} / \tau_{\rm lib}$. For a given disc model (density, temperature, alpha viscosity), the amplitude of the dynamical corotation torque thus depends on the planet's mass and orbital distance. To assess how the dynamical corotation torque impacts type I migration, we could have carried out a number of 1D simulations, like those of the previous section, with varying planet mass and initial location, possibly by modelling the mass growth of the planet. For the present work, we adopt a simpler, more pragmatic approach by providing a simple estimate for how much the dynamical corotation torque is expected to reduce the classical type I migration torque, comprised of only the Lindblad and static corotation torques. 

We adopt a locally isothermal, low-viscosity disc model with shallow radial profiles of density and temperature, such that the dynamical corotation torque slows down type I migration. We shall assume the migration timescale is long compared to the libration timescale (i.e., $\dot{r}_\mathrm{P} \ll 3\Omega_\mathrm{P}x_\mathrm{s}^2/\{4\pi r_\mathrm{P}\}$) so that Eq.~(\ref{eq:dyn_crt_analytic}) can be recast as \citep{paardekooper_dynamical_2014}
\begin{equation}\label{eq:dyn_crt_slow}
    \Gamma_\mathrm{c,dyn} = 2\pi \left( 1-\frac{\Ivlib}{\Ivcross} \right) \Sigma_\mathrm{P} r_\mathrm{P}^2 x_\mathrm{s}\Omega_\mathrm{P} \dot{r}_\mathrm{P}.
\end{equation}
Writing the total disc torque $\Gamma$ as $\Gamma_\mathrm{lin}+\Gamma_\mathrm{c,static}+\Gamma_\mathrm{c,dyn}$ and as $ \frac{1}{2}M_\mathrm{P} r_\mathrm{P} \Omega_\mathrm{P} \dot{r}_\mathrm{P}$ (neglecting the time evolution of the planet's mass), one arrives at
\begin{equation}\label{eq:total_torque}
    \Gamma = \frac{\Gamma_\mathrm{lin}+\Gamma_\mathrm{c,static}}{1-\frac{M_\mathrm{HS}}{M_\mathrm{P}}\left(1 - \frac{\Ivlib}{\Ivcross}\right)},
\end{equation}
where $M_\mathrm{HS}=4\pi r_\mathrm{P}x_\mathrm{s}\Sigma_\mathrm{P}$ is the mass inside the HS region. The total disc torque on the planet can thus be expressed as a simple correction to the classical type I torque,
\begin{equation}\label{eq:reduction_factor}
    \Gamma = f_\mathrm{red} \times \left( \Gamma_\mathrm{lin} + \Gamma_\mathrm{c,static} \right) 
\end{equation}
with $f_\mathrm{red} < 1$ a reduction factor that reads
\begin{equation}
    f_\mathrm{red} = \left[ 1-\frac{M_\mathrm{HS}}{M_\mathrm{P}}\left(1 - \frac{\Ivlib}{\Ivcross} \right) \right]^{-1}.
    \label{eq:fred_explicit}
\end{equation}
When $\Gamma_\mathrm{c,dyn} \rightarrow 0$, $f_\mathrm{red} \rightarrow 1$, whereas when $\Gamma_\mathrm{c,dyn}$ is large, $f_\mathrm{red} \ll 1$. Again, for a given disc model, $f_\mathrm{red}$ depends on the planet's mass and orbital radius.

We evaluate the reduction factor $f_\mathrm{red}$ for disc models similar to those used in global models of planet formation and evolution:
\begin{eqnarray}
    \Sigma(r) &=& \Sigma_{\mathrm{5\,au}} \times (r/\mathrm{5\,au})^{-\beta}     \\
    T(r) &=& T_{\mathrm{1\,au}} \times (r/\mathrm{1\,au})^{-1/2},
\end{eqnarray}
with $\Sigma_{\mathrm{5\,au}}=300\,\mathrm{g}\,\mathrm{cm^{-2}}$, $\beta \in [1/2,1]$, and $T_{\mathrm{1\,au}}=280\,\mathrm{K}$. Our temperature profile (decreasing as $r^{-1/2}$) mimics that of a passively irradiated disc. To compute the dynamical corotation torque, we first compute the classical type I migration torque, which we dub the static torque, via the torques formulae in \cite{Paardekooper2011}, and use the migration rate resulting from the sole static torque to compute the dynamical corotation torque through Eq.~(\ref{eq:dyn_crt_slow}). Like in Section~\ref{sec:1D_simulations}, $\Ivcross$ is obtained with Eq.~(\ref{eq:Ivcross}) and $\Ivlib$ via Eqs.~(\ref{eq:Ivlib}) and~(\ref{eq:final_memory_timescale}), assuming the planet has migrated sufficiently far for it to loose its memory of the initial location (i.e. $r_\mathrm{P}-r_0< \dot{r}_\mathrm{P} \cdot \tau_\mathrm{memory}$). We then iterate, using the migration rate associated to the total torque to compute again the dynamical corotation torque, until the value converges. The reduction factor is finally calculated as $1 + \Gamma_\mathrm{c,dyn} / (\Gamma_\mathrm{lin} +  \Gamma_\mathrm{c,static})$. We note that, although our memory timescale formulation for the dynamical corotation has been worked out for uniform temperature profiles, it also reproduces well the results of 2D hydrodynamical simulations with non-uniform temperature profile (locally isothermal discs), as will be shown in Section~\ref{sec:disc_2Dsimulations}.

\begin{figure*}
    \centering
    \includegraphics[width=\hsize]{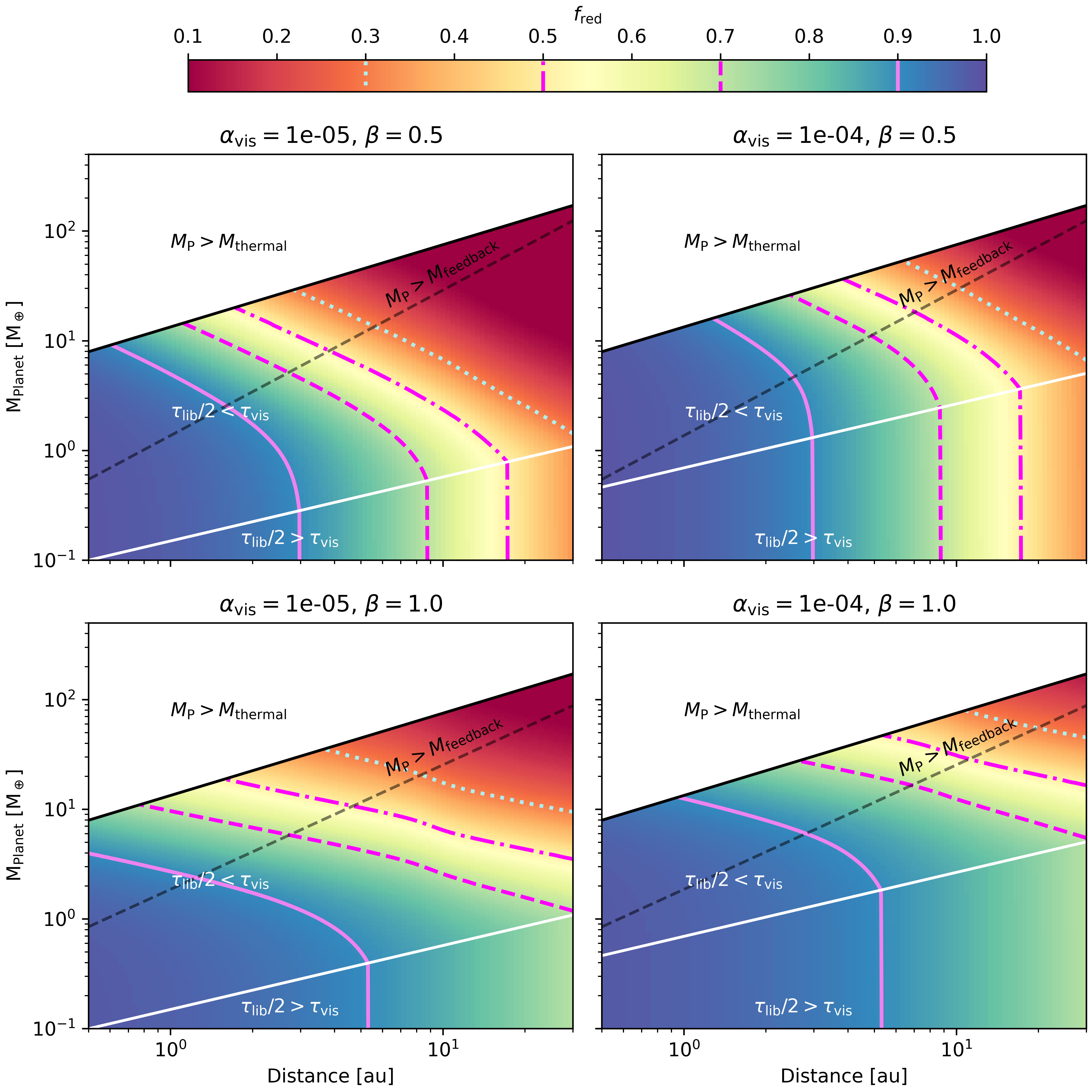}
    \caption{Maps showing the reduction factor $f_\mathrm{red}$ defined by Eq.~(\ref{eq:reduction_factor}), as a function of planet mass and orbital distance. The quantity $f_\mathrm{red}$ quantifies how much the classical type I migration torque (sum of the Lindblad and static corotation torques) is reduced by the dynamical corotation torque: the latter is negligible where $f_\mathrm{red} \rightarrow 1$ and strong where $f_\mathrm{red} \ll 1$. Maps are obtained for  four disc models that differ by the choice of the alpha turbulent viscosity ($\alpha_{\rm vis}$) and the slope of the density profile ($-\beta$). Contour lines are plotted for $f_\mathrm{red}\in[0.3,0.5,0.7,0.9]$ for better visualisation. The white solid line marks the transition where $\tau_\mathrm{lib} / 2 = \tau_\mathrm{vis}$ (see text). The dashed black line marks the feedback mass (see text). Regions where the planet mass exceeds the thermal mass (equivalent to $q_{\rm P} / h^3(r_{\rm P}) > 1$) are excluded as there we expect substantial gap opening to occur in low-viscosity discs.}
    \label{fig:dyncrt_reduction_factor}
\end{figure*}

Figure~\ref{fig:dyncrt_reduction_factor} displays the reduction factor $f_\mathrm{red}$ as a function of planet mass and orbital distance for four disc models with $\alpha_\mathrm{vis}\in[10^{-5},10^{-4}]$ and $\beta \in [1/2,1]$. Each panel shows two regimes:
\begin{itemize}
\item[(i)] where $\tau_\mathrm{lib}/2 > \tau_\mathrm{vis}$ (below the white solid line): $f_\mathrm{red}$ shows no dependence on alpha turbulent viscosity and planet mass, but it decreases with radial distance. This can be explained as follows: in this regime, viscous diffusion implies that $\Ivlib \approx \Iv(r_{\mathrm P})$, and from Eq.~(\ref{eq:Ivcross}) $\Ivlib / \Ivcross \approx 1 + (x_{\rm s}/r_{\rm p})(3/2-\beta)$. Eq.~(\ref{eq:fred_explicit}) then yields
\begin{equation}
f_\mathrm{red} \approx \frac{1}{1 + 4\pi\Sigma x^2_{\rm s} \left( \frac{3}{2}-\beta \right) / M_{\rm P} }, 
\end{equation}
which is indeed independent of $\alpha_{\rm vis}$ and $M_{\rm P} $ (since $x_{\rm s}\propto \sqrt{M_{\rm P}}$) and decreases with radial distance for our radial profiles of density and temperature.
\smallskip
\item[(ii)] where $\tau_\mathrm{lib}/2 < \tau_\mathrm{vis}$ (above the white solid line): here, a memory effect sets in for the librating vortensity and we see that $f_\mathrm{red}$ now decreases with planet mass, but increases with alpha viscosity. We interpret this as a consequence of $\Ivlib$ scaling as $(1 - \tau_{\rm memory}/\tau_{\rm mig})^{3/2-\beta}$ according to Eq.~(\ref{eq:Ivlib}) with $\tau_{\rm memory} \propto \tau_{\rm vis}$ (Eq.~\ref{eq:final_memory_timescale}). Since $\tau_{\rm mig} \propto 1/M_{\rm P}$ and $\tau_{\rm vis} \propto M^2_{\rm P}/\alpha_{\rm vis}$, increasing the planet mass or decreasing the alpha viscosity leads to a higher $\Ivlib / \Ivcross$ ratio, a stronger dynamical corotation torque, and thus a smaller reduction factor $f_\mathrm{red}$.
\end{itemize}

The panels clearly show that, overall, the reduction of the classical type I migration torque by the dynamical corotation torque is stronger at lower $\alpha_{\rm vis}$ and lower $\beta$ (as already seen in the previous sections), but also for higher planet mass and larger orbital distances. This is probably best seen by looking at the position of the dash-dotted curve throughout the panels, which marks where the dynamical corotation torque reduces the classical type I migration torque by a factor 2.

Classical type I migration occurs so long as the planet wakes do not significantly perturb the disc's background density and/or temperature profiles. The threshold mass of a planet above which its wakes start altering the background disc depends on the sound speed, the turbulent viscosity, and the relative velocity between the disc and planet (it regulates how frequently given fluid parcels get shocked through the wakes; see, e.g., \citealp{wafflard-fernandez_intermittent_2020}). Different indicators can be used to estimate this threshold mass. One is the thermal mass, which is the mass from which the planet's Hill radius exceeds a fraction of the pressure scale height, which we simply write $M_\mathrm{thermal}=h^3 M_\star$. Another is the feedback mass, which is given by $M_\mathrm{feedback} \approx 3.8 \times \left(Q/h\right)^{-5/13} \times 2/3 M_\mathrm{thermal}$ \citep[Eq. 53 in][]{rafikov_planet_2002}. It corresponds to the mass from which migration becomes impacted by the opening of a small gap around the planet orbit, which often comes with the formation of vortices due to the Rossby-Wave instability setting at the gap edges, in particular in low-viscosity discs \citep{McNally2019,ziampras_migration_2024,ziampras_halting_2025}. Note that the feedback mass expression has been derived for inviscid discs and discards the disc-planet relative drift (due for instance to planet migration). The solid and dashed black curves in Figure~\ref{fig:dyncrt_reduction_factor} show where the planet mass exceeds the thermal mass and the feedback mass, respectively.

The outcome of global models of planet formation and evolution and planetary population syntheses using such models are sensitive to the planet's growth and migration timescales \citep{BurnMordasini24}. It is a well-known problem for low-mass planets, which pose the challenge of growing faster than they migrate inward. This situation is also relevant to our study as the dynamical corotation torque comes into play once the planet has grown to several Earth masses and/or remains at fairly large orbital separations ($\gtrsim$ 10 au). A detailed investigation of the interplay between growth and migration via global models including our formulation of the dynamical corotation torque is the subject of a future study.

\section{Discussion}
\label{sec:discussion}

\subsection{On the 2D simulations}
\label{sec:disc_2Dsimulations}

In this study we have presented the results of 2D hydrodynamical simulations for a specific physical model and numerical setup, and we now discuss how these impact our results.
\\
\par \noindent{\it Effects of grid resolution -- } The 2D simulations presented in Section~\ref{sec:2D_simulations} have $(N_\mathrm{r},N_\phi)=(782,2346)$, with grid cells spaced linearly or logarithmically along the radial direction. The full width of the planet's HS region is initially resolved by about 10 cells, which is sufficient to capture the effects of vortensity mixing in the HS region within reasonable runtime. However, a finite grid resolution inevitably leads to numerical diffusion effects, which can be seen as an equivalent numerical viscosity. To see how it impacts our results, we ran additional test simulations with $Q_0=8$, $\beta=1$ for $\alpha_\mathrm{vis} = \{0, 10^{-7}, 10^{-6}$\}. The simulations with $\alpha_\mathrm{vis}=10^{-7}$ and $\alpha_\mathrm{vis}=0$ show identical results that both deviate from those of the run with $\alpha_\mathrm{vis}=10^{-6}$. We thus conclude that for our setup, the effects of numerical diffusion can be seen as an equivalent alpha viscosity whose magnitude is between $10^{-7}$ and $10^{-6}$. Furthermore, as the planet migrates inwards, its HS region gets narrower ($x_\mathrm{S} \propto r_{\rm P}^{1-f/2}$) and less well resolved for a grid with linear radial spacing. Again, additional test simulations have shown that a grid with logarithmic radial spacing had to be used in our runs where the planet comes close to the grid's inner wave-damping zone (which is the case of our default $Q_0 = 8$ runs for $\beta = \{3/2,2\}$), more especially at low viscosities. This explains the grid choices made in our set of simulations (see Table~\ref{tab:fargo_simlist}).
\\
\par
\begin{figure*}[]
    \centering
    \includegraphics[trim={1cm 2cm 2cm 0},clip,width=\hsize]{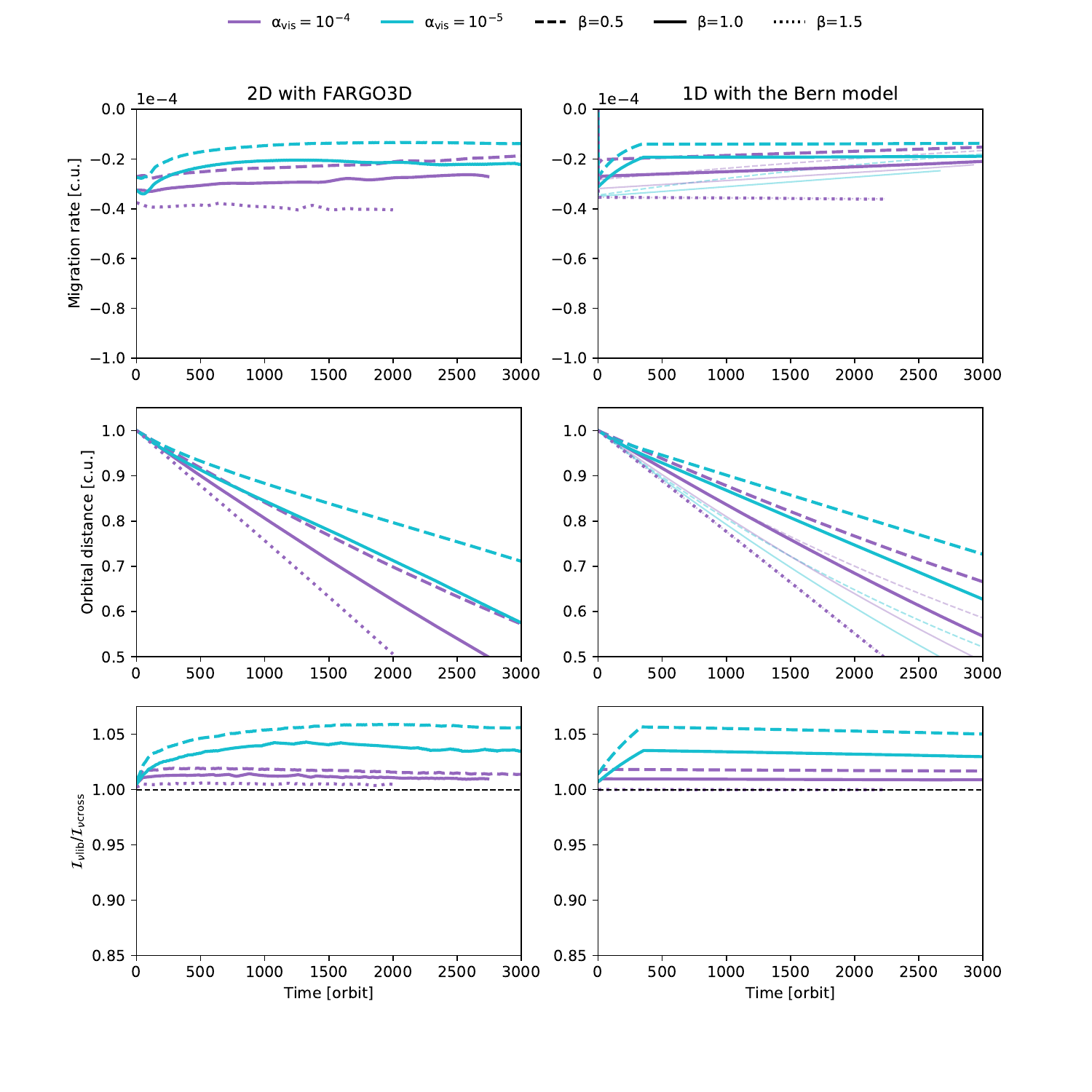}
    \caption{Same as Figure~\ref{fig:simulations_TQ8_migrate_orbital_evo}, for additional simulations with a background temperature profile decreasing as $1/r$, demonstrating the ability of our simple 1D model to also reproduce non-isothermal simulation results.}
    \label{fig:non-isothermal_calculations}
\end{figure*}

 \noindent{\it Non-isothermal simulations -- } Our 2D simulations have assessed the evolution of the disc's vortensity in the planet's HS region in globally isothermal discs. As already stated in Section~\ref{sec:setup_2D_simulations}, this choice was motivated by avoiding strong vortensity perturbations close to the planet's location, and thus aimed to facilitate the (initial) estimation of $\Ivlib$ and $\Ivcross$ from the simulations results. However, protoplanetary discs are not thought to have a uniformly radial temperature profile, and we have therefore carried out a set of additional 2D and 1D simulations with a temperature profile $T(r)\propto r^{-1}$ (the disc's flaring index thus cancels out), for $\alpha_\mathrm{vis}\in\{10^{-5},10^{-4}\}$ and $\beta\in\{1/2,1,3/2\}$ (all 2D simulations use a grid with logarithmic radial spacing). The results of these simulations are displayed in Figure \ref{fig:non-isothermal_calculations}, and show an overall good agreement between the 1D and 2D runs, both for the $\Ivlib / \Ivcross$ ratios and the migration rates. It indicates that our formulation of the dynamical corotation torque is not restricted to uniform temperature profiles. Still, we stress that the present study assumes discs described by a locally isothermal equation of state, in which the disc's entropy gradient has no impact on the static and dynamical corotation torques. Future work will aim to examine how to extend our new formulation for the dynamical corotation torque to radiative disc models (discs described by an energy equation with heating and cooling source terms). 

\subsection{An alternative 1D model for the dynamical corotation torque}
\label{sec:alternative_1D_model}

As recalled in Section~\ref{sec:fund_analytical_theory}, the dynamical corotation torque depends on the inverse vortensities of the librating flow ($\Ivlib$) and the orbit-crossing flow ($\Ivcross$). Its inclusion in 1D simulations of planet formation and evolution necessitates expressions for $\Ivlib$ and $\Ivcross$, which is relatively straightforward for $\Ivcross$ (Eq.~\ref{eq:Ivcross}) but much less straightforward for $\Ivlib$ since the latter depends on the planet's migration history. We have shown that the time evolution of $\Ivlib$ in 2D simulations of disc-planet interactions is well reproduced by a memory timescale argument, and the results of 2D simulations were used to estimate the memory timescale.

During the course of this study, we have come out with another simple, empirical model to evaluate $\Ivlib$ and $\Ivcross$, which can be expressed as follows:
\begin{equation}
	\omega_{\mathrm{lib}}(t) = \frac{\tau_\mathrm{vis}\omega_0(r_{\rm P}(t=0)) + t\omega(r_\mathrm{P}(t))}{\tau_\mathrm{vis} + t},
	\label{eq:omegalib_clem}
\end{equation}
\begin{equation}
	\omega_{\mathrm{cross}}(t) = \omega(r_\mathrm{P}(t) - x_{\rm s}(t)),
	\label{eq:omegacross_clem}
\end{equation}
where $\omega_0$ and $\omega$ denote the initial and instantaneous vortensity profiles, $\tau_{\rm vis}$ is the viscous timescale across the planet's HS region (see Section~\ref{sec:memory_sect2}), and where $\Ivlib(t) = \omega^{-1}_{\mathrm{lib}}(t)$ and $\Ivcross(t) = \omega^{-1}_{\mathrm{cross}}(t)$. Eq.~(\ref{eq:omegacross_clem}) is simply the same as Eq.~(\ref{eq:Ivcross}). As for Eq.~(\ref{eq:omegalib_clem}), the rationale behind the linear combination is as follows. When $t \ll \tau_{\rm vis}$ (for instance for very low turbulent viscosities), $\omega_{\mathrm{lib}}(t) \approx \omega_0(r_{\rm P}(t=0))$: the librating flow keeps its initial vortensity, as expected when vortensity advection dominates over vortensity diffusion. And when $t \gg \tau_{\rm vis}$, $\omega_{\mathrm{lib}}(t) \approx \omega(r_\mathrm{P}(t))$: the vortensity of the librating flow adjusts to the instantaneous local vortensity, as expected when vortensity diffusion dominates over vortensity advection. Eqs.~(\ref{eq:omegalib_clem}) and (\ref{eq:omegacross_clem}) yields the following limits for $\Ivlib / \Ivcross$: 
\begin{eqnarray}
\label{eq:clem_limit1}
\frac{\Ivlib}{\Ivcross} &\approx& 1 - \frac{x_{\rm s}(t)}{r_{\rm P}(t)}\frac{d\log\omega}{d\log r}\Bigg|_{r_\mathrm{P}(t)}  \;\; {\rm for}\;\tau_{\rm vis} \ll t\\
&\approx& \frac{\omega(r_{\rm P}(t)-x_{\rm s}(t))}{\omega_0(r_{\rm P}(t=0))} \;\; {\rm for}\;\tau_{\rm vis} \gg t.
\label{eq:clem_limit2}
\end{eqnarray}
In the first limit (Eq.~\ref{eq:clem_limit1}), $\Ivlib / \Ivcross$ is entirely set by the background vortensity gradient, while the second limit (Eq.~\ref{eq:clem_limit2}) corresponds to the case where the librating flow has infinite vortensity memory.

\begin{figure*}[]
    \centering
    \includegraphics[trim={1cm 1.5cm 2cm 0},clip,width=\hsize]{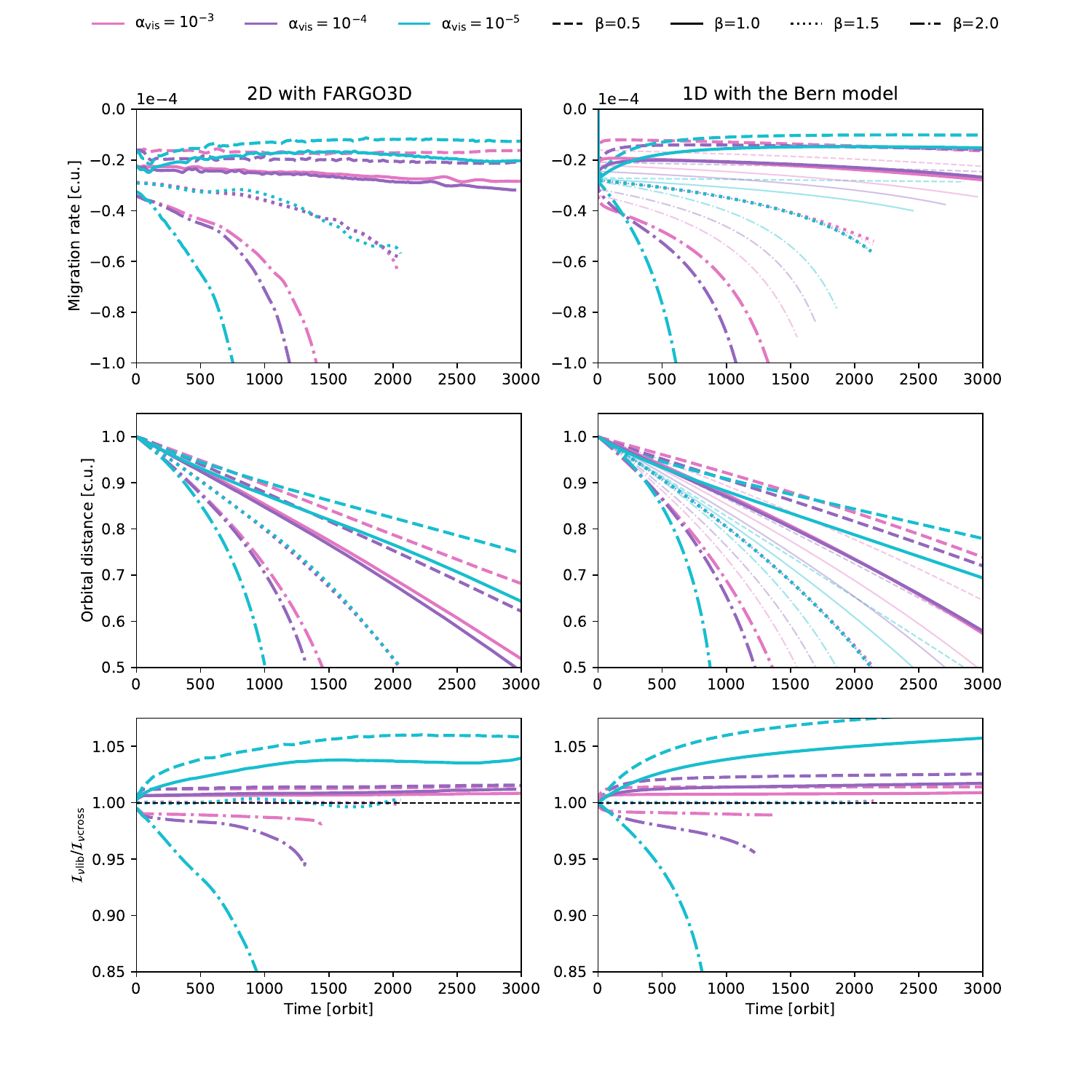}
    \caption{Same as Figure~\ref{fig:simulations_TQ8_migrate_orbital_evo}, except that 1D simulations use the alternative formulation for the dynamical corotation torque of Section~\ref{sec:alternative_1D_model}. It shows remarkable agreement for the runaway migration ($\beta>3/2$) and good overall agreement for the slow-down migration simulations. For an easier direct comparison of the most relevant simulations, see right column in Figure \ref{fig:direct_comparison} in Appendix \ref{app:direct_compar}.}
    \label{fig:simulations_TQ8_migrate_evo_Clement}
\end{figure*}

Figure \ref{fig:simulations_TQ8_migrate_evo_Clement} compares the results of the 2D simulations presented in Section~\ref{sec:results} with those of 1D simulations using the same setup as in Section~\ref{sec:setup_1D} but with $\Ivlib$ and $\Ivcross$ computed from Eqs.~(\ref{eq:omegalib_clem}) and~(\ref{eq:omegacross_clem}). We see that the alternative model proposed in this section also works very well, overall. We see that for $\beta < 3/2$, the alternative model better reproduces the initial increase in $\Ivlib / \Ivcross$ but overestimates a bit its stationary value, thereby predicting slightly slower inward migration than with our fiducial memory model. The 1D simulations for $\beta > 3/2$ show quite remarkable agreement with the 2D simulations.

\subsection{Comparison to previous work}
\label{sec:comp_previous_work}
A simple model for evaluating $\Ivlib / \Ivcross$ was derived in \cite{paardekooper_dynamical_2014}, based on the evolution toward a steady state of the vortensity in the planet's HS region due to viscous diffusion and planet migration (see Equation 28 of \citealp{paardekooper_dynamical_2014}):
\begin{equation}
    1 - \frac{\Ivlib}{\Ivcross} = \left( \frac{3}{2} - \beta \right) \frac{x_\mathrm{s}^2}{6r_\mathrm{P}\nu} \dot{r}_\mathrm{P},
\end{equation}
which can be recast as
\begin{equation}
    \frac{\Ivlib}{\Ivcross} = 1 - \left( \frac{3}{2} - \beta \right) \frac{\tau_\mathrm{vis}}{6\tau_\mathrm{mig}}.
    \label{eq:ratio_sj}
\end{equation}
When $\tau_{\rm lib} \ll \tau_{\rm vis} \ll |\tau_{\rm mig}|$, the memory timescale in Eq.~(\ref{eq:final_memory_timescale}) is $\approx 0.4\tau_{\rm vis} = 2\tau_{\rm vis}/5$, and Eq.~(\ref{eq:Ivlib}) yields
\begin{equation}
    \frac{\Ivlib}{\Ivcross} \approx 1 - \left( \frac{3}{2} - \beta \right) \frac{2\tau_\mathrm{vis}}{5\tau_\mathrm{mig}},
    \label{eq:ratio_jesse}
\end{equation}
which is similar to the original model of \cite{paardekooper_dynamical_2014}, except for the constant factor in front of $\tau_{\rm vis}/\tau_{\rm mig}$. We have tested this model by running again 1D simulations with the same setup as in Section~\ref{sec:setup_1D}, using for the dynamical corotation torque the expression in Eq.~(\ref{eq:dyn_crt_slow}), which is valid when $\tau_{\rm lib} \ll |\tau_{\rm mig}|$, and using for $\Ivlib / \Ivcross$ the expression in Eq.~(\ref{eq:ratio_sj}).

\begin{figure*}[]
    \centering
    \includegraphics[trim={1cm 1.5cm 2cm 0},clip,width=\hsize]{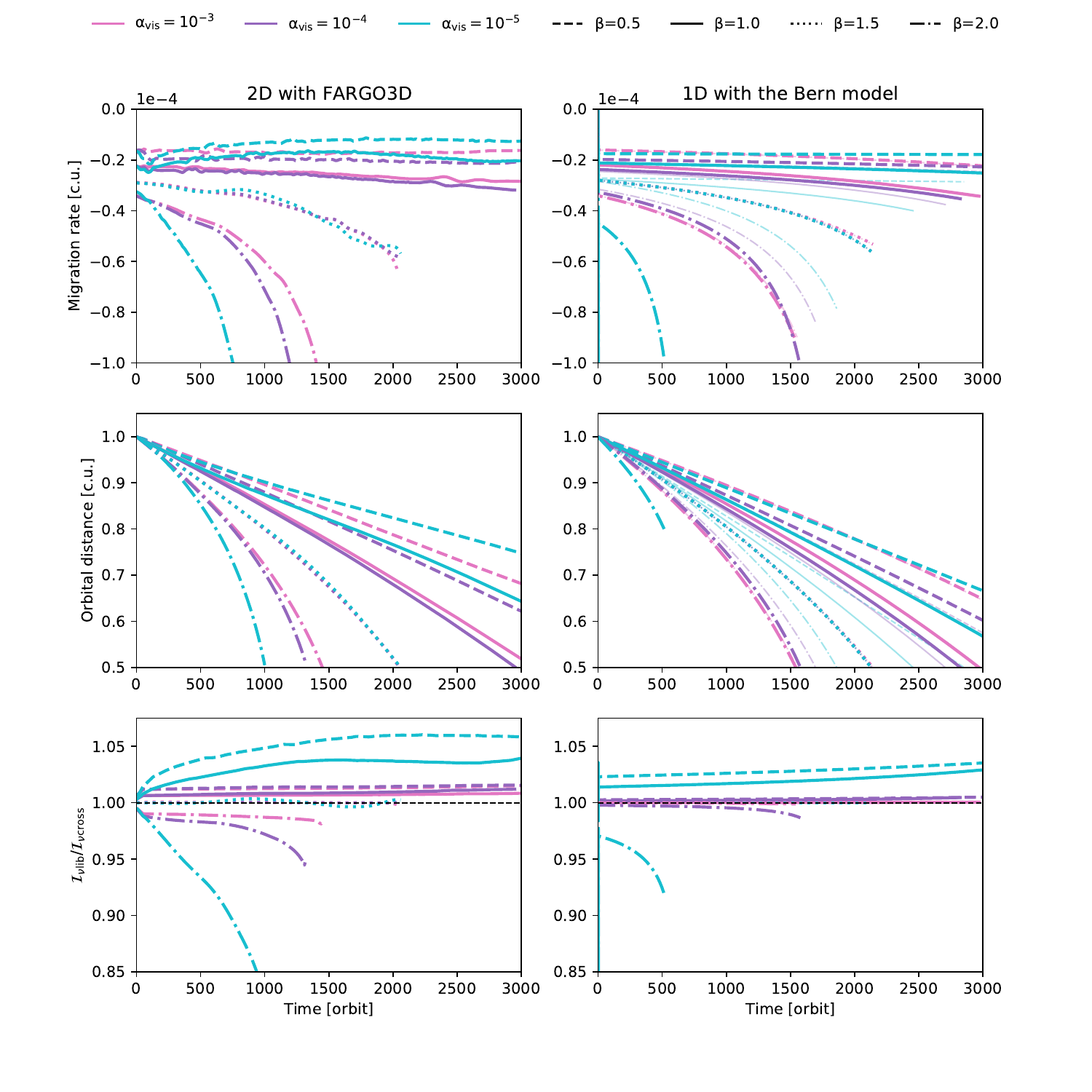}
    \caption{Same as Figure~\ref{fig:simulations_TQ8_migrate_orbital_evo}, except that 1D simulations use the original formulation for the dynamical corotation torque of \citet{paardekooper_dynamical_2014}, which is recalled in Section~\ref{sec:comp_previous_work}. Note that this steady state approach is not able to reproduce the initial build up of vortensity ratio and the final values are generally under predicted.}
    \label{fig:simulations_TQ8_migrate_evo_Paardekooper2014}
\end{figure*}

Figure \ref{fig:simulations_TQ8_migrate_evo_Paardekooper2014} compares the results of these 1D simulations with those of the 2D simulations presented in Section~\ref{sec:results}. We see that the 1D simulations underestimate the ratio $\Ivlib / \Ivcross$ and therefore predict faster inward migration than observed in our 2D simulations. This is maybe related, at least partly, to the constant factor in front of $\tau_{\rm vis}/\tau_{\rm mig}$ (1/6 vs. 2/5 between Eqs.~\ref{eq:ratio_sj} and~\ref{eq:ratio_jesse}). Also, since Eq.~(\ref{eq:ratio_sj}) was derived as a stationary solution, it cannot reproduce the gradual increase in $\Ivlib / \Ivcross$ before reaching a quasi-steady state. Finally, we see that the results of the 1D runs significantly depart from those of the 2D runs for $\beta>3/2$. This is not surprising since the above formulation for $\Ivlib / \Ivcross$ assumes a steady migration rate, which does not apply to the runaway migration regime. We briefly note that, due to a fast increase in the migration rate for $\alpha_\mathrm{vis}=10^{-5}$ and $\beta=2$, our 1D simulation showed unstable behaviour and was stopped after $\sim 500$ orbits. To our knowledge, the above results constitute the first systematic benchmark of the dynamical corotation torque model of \cite{paardekooper_dynamical_2014} against hydrodynamical simulations of disc-planet interactions.

Since the study of \citet{paardekooper_dynamical_2014}, the growth of interest for understanding planet migration and evolution in low-viscosity discs has brought up a number of works on the dynamical corotation torque via 2D and 3D (magneto-)hydrodynamical simulations, and simple 1D simulations. Here we briefly summarise their results. A first attempt at including the effects of the dynamical corotation torque in 1D planet formation models was done by \cite{ogihara_effects_2017} to study the effects of magnetised disc winds on type I migration. They used the dynamical corotation torque expression of \cite{paardekooper_dynamical_2014} with $\Ivlib / \Ivcross$ given by Eq.~(\ref{eq:ratio_sj}), but where $\tau_{\rm mig}$ is the characteristic timescale for the relative migration between the planet and disc. They predicted that super-Earths would have a bimodal distribution of semi-major axes due to how winds affect the disc structure and the resulting type I migration.  \cite{Ndugu2021} investigated the dynamical corotation torque by assuming conservation of vortensity in the horseshoe region but calculating the evolution of the dynamical corotation torque by evolving the mass of the horseshoe region (including effects of gap opening, viscous spreading and planet accretion). These authors concluded that planet accretion from the horseshoe region and the dynamical corotation torque can efficiently slow down type I migration. In \cite{mcnally_low_2017} and \cite{mcnally_low-mass_2018}, the effect of a net radial in- or outflow induced by Maxwell stresses in low-viscosity discs was also investigated. They found a dynamical corotation torque arising from the flow-induced asymmetric distortion of the corotation region, leading to four distinct migration regimes depending on the setup (i.e. locked migration to the inward or outward radial gas flow and outward runaway migration). They found that runaway outward migration saturates to a steady speed and provided an analytical model. In \cite{mcnally_low-mass_2020}, it was shown that a wind-driven laminar accretion flow through the disc surface layer does not impact much the torques exerted onto embedded planets. However, they also showed that the dynamical corotation torque can change behaviour dramatically when going from 2D to 3D, switching sign and leading to fast inward migration due to the dissipation of buoyancy waves \citep[see also][]{zhu_planet-disk_2012}. 
Further 3D inviscid radiation hydrodynamic simulations showed that radiative cooling and diffusion can damp buoyancy waves more or less efficiently depending on the disc's local physical properties \citep{yun_effects_2022, ziampras_buoyancy_2024}, which affects the sign and magnitude of the dynamical corotation torque. All the aforementioned works show that a complete picture of the dynamical corotation torque has not yet emerged and that there is still much room to improve its formulation for global models of planet formation and evolution, which can serve via population syntheses to compare the different models with observations.

\section{Summary and conclusions}
The dynamical corotation torque can greatly impact the migration of low-mass planets in their protoplanetary disc, particularly for low-viscosity wind-driven discs. Yet, it is often discarded in global models of planet formation and evolution, which rely on 1D simulations. Its calculation requires indeed analytical prescriptions for the vortensity of the librating and orbit-crossing flows in the planet's horseshoe region, which for the librating flow depends on the migration history of the planet. In this work, we have proposed two simple prescriptions for the time evolution of the librating vortensity.

Our main prescription is based on a memory effect (Section~\ref{sec:memory_sect2}), which features a memory timescale for the librating flow to forget about its initial vortensity and adjust to that of the ambient gas, after which the ratio of inverse vortensities between the librating and orbit-crossing flows becomes stationary. 2D hydrodynamical simulations assuming locally isothermal discs were used to estimate this memory timescale, which is a fraction of the viscous timescale across the horseshoe region. An alternative prescription for the inverse vortensities of the librating and orbit-crossing flows is proposed in Section~\ref{sec:alternative_1D_model}.

1D simulations including aforementioned prescriptions were carried out and compared to 2D hydrodynamical simulations for locally isothermal discs. The planet's orbital evolutions in the 1D and 2D simulations are in very good agreement for a wide parameter space, whether the dynamical corotation torque slows down inward migration or runs it away. For typical shallow density and temperature profiles in discs, we have provided maps showing how much the dynamical corotation torque reduces the classical type I migration torque (sum of the Lindblad and static corotation torques) as a function of planet mass and orbital distance (Fig.~\ref{fig:dyncrt_reduction_factor} in Section~\ref{sec:influence_on_planet_formation}). This reduction is larger at higher planet mass and larger orbital distance. For a young disc with surface density profile in $r^{-1/2}$ and $\alpha_{\rm vis} = 10^{-4}$, it is by a factor $\approx$50\% for a 10 Earth-mass planet at 10 au.

In summary, the dynamical corotation torque should be taken into account in global models of planet formation and evolution for discs with low turbulent viscosity ($\alpha_\mathrm{vis}\lesssim 10^{-4}$). The presented framework of using hydrodynamical simulations of disc-planet interactions to inform 1D analytical prescriptions for the vortensity in the planet's horseshoe region will help assess, in the future, the impact of other physical processes on the dynamical corotation torque, like for instance vortensity production due to baroclinic forcing \citep{ziampras_buoyancy_2024,ziampras_migration_2024}.

\begin{acknowledgements}
We thank Oliver Schib, Richard Nelson, Aurélien Crida, Alessandro Morbidelli, Nicolas Kaufmann, and Andrin Kessler for stimulating discussions on this work. We further thank Alessandro Morbidelli for a thorough reading of a first version of this manuscript. We also thank the anonymous referee for their comments that helped improve the manuscript. J.W. and C.M. acknowledge the support from the Swiss National Science Foundation under grant 200021\_204847 “PlanetsInTime”. Part of this work has been carried out within the framework of the NCCR PlanetS supported by the Swiss National Science Foundation under grants 51NF40\_182901 and 51NF40\_205606. Calculations were performed on the Horus cluster of the Division of Space Research and Planetary Sciences at the University of Bern. Part of the numerical simulations were performed on the CALMIP Supercomputing Centre of the University of Toulouse.
\end{acknowledgements}

\bibliographystyle{bibtex/aa.bst}
\bibliography{references.bib}
\onecolumn
\begin{appendix}
    
    \section{Additional simulations with higher aspect ratio} \label{app:high_aspect_ratio}
\begin{figure*}[h!]
        \centering
        \includegraphics[trim={1cm 1cm 2cm 0},clip,width=0.9\hsize]{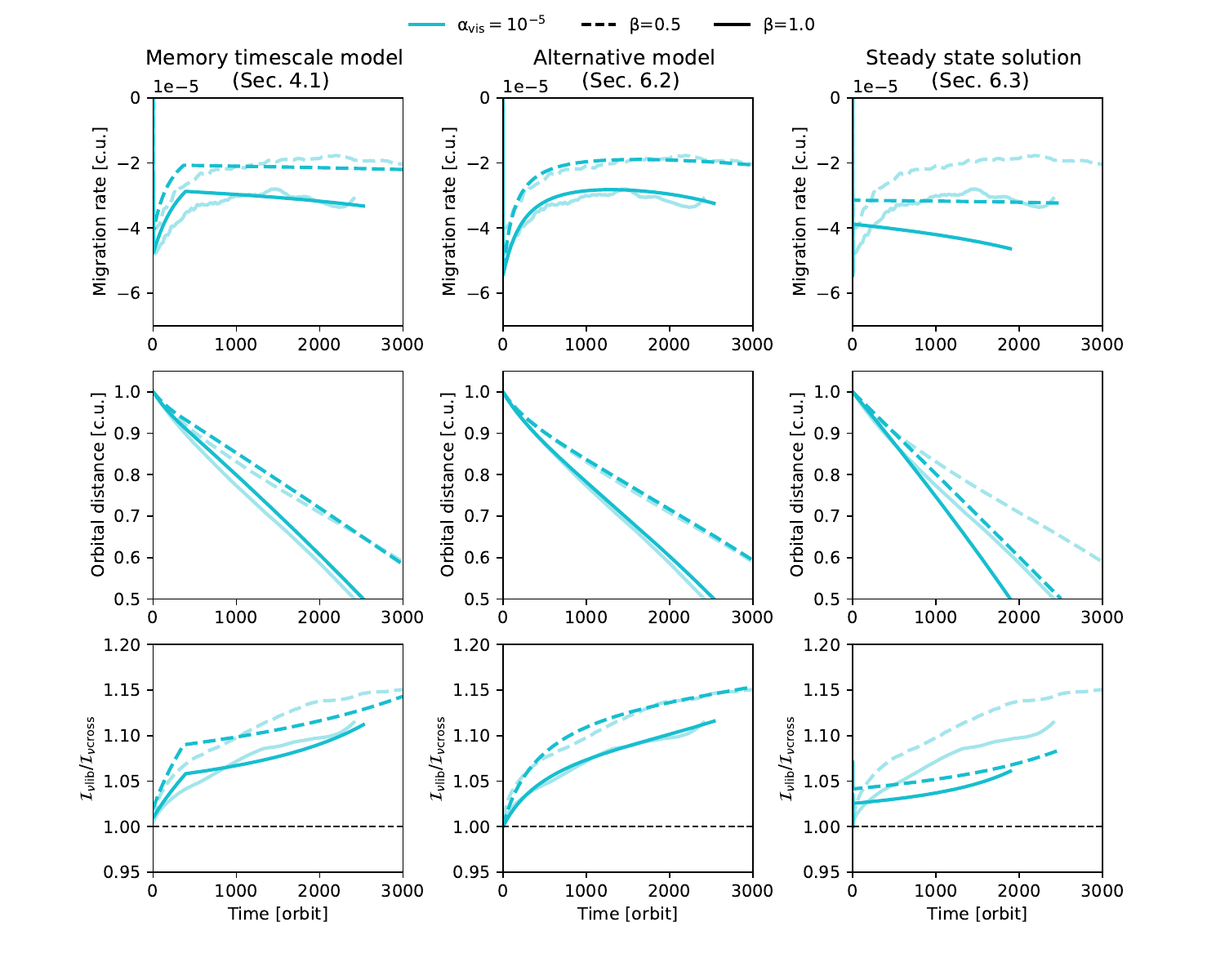}
        \caption{Direct comparison of 2D and 1D simulations results for $h_0=0.07$, $q_{\rm p}=2.7\times 10^{-5}$ and $Q_0=8$. Rows are the same as in Fig. \ref{fig:simulations_TQ8_migrate_orbital_evo}. Columns correspond to different approaches for calculating the inverse vortensities $\Ivlib$ and $\Ivcross$: the memory timescale model (Section~\ref{sec:setup_1D}, left), the alternative model (Section~\ref{sec:alternative_1D_model}, middle) and the steady state solution (Section~\ref{sec:comp_previous_work}, right). Opaque lines are for 1D simulations, transparent lines for 2D simulations.}
        \label{fig:direct_comparison_ar7em2}
    \end{figure*}

    In the main text, we have presented results of simulations for inwardly migrating low-mass planets starting at a disc's aspect ratio $h_0=0.05$. Since our fiducial disc model assumes $h(r) \propto r^{1/2}$, our simulations automatically tested the impact of having a smaller aspect ratio through migration (down to $h(0.5r_0) \sim 3.5\%$). In order to further test our 1D models of the dynamical corotation torque against larger aspect ratios, we have carried out additional simulations for $h_0=0.07$. We have increased the planet mass to maintain the same $q_{\rm p}/h_0^3$ ratio as in our fiducial model, such that the density perturbations associated to the planet wakes reach very similar values. We have also increased $\Sigma_0$ to retain $Q_0=8$. The initial migration rate, which scales as $q_{\rm p} / \{h_0Q_0\}$, is about twice as large as in our fiducial model for the same $Q_0$. These additional simulations used $\alpha_\mathrm{vis}=10^{-5}$ and $\beta=\{\frac{1}{2},1\}$. Results are shown in Figure~\ref{fig:direct_comparison_ar7em2}.

    The memory timescale model (Section~\ref{sec:setup_1D}) predicts a slightly different time evolution for the inverse vortensity ratio (see bottom-left panel), but the magnitude is well reproduced and therefore so is the migration rate. The alternative model of Section~\ref{sec:alternative_1D_model} reproduces nearly perfectly the time evolution of the inverse vortensity ratio, the migration rate and the planet's orbital distance. The steady-state solution model discussed in Section~\ref{sec:comp_previous_work} underpredicts the inverse vortensity ratio, leading to too high migration rates.
    
    We finally note that these simulations all show a constantly increasing inverse vortensity ratio.
    In both our memory timescale model and the steady-state solution model, $\Ivlib / \Ivcross - 1$ scales with $\tau_\mathrm{vis}/\tau_\mathrm{mig}$ (see Eqs.~\ref{eq:ratio_sj} and~\ref{eq:ratio_jesse}; these are valid for $\tau_\mathrm{lib} \ll \tau_\mathrm{vis} \ll |\tau_\mathrm{mig}|$, and the latter inequality is marginally fulfilled in the simulations presented here). Since $\tau_{\rm vis} \equiv q_{\rm p} / \{\alpha h^3(r_{\rm p})\} \times \Omega^{-1} \propto r^{3/2-3f}_{\rm p}$, $\tau_{\rm vis}$ remains stationary for $f=1/2$. When considering only the Lindblad torque for simplicity, $|\tau_{\rm mig}| \propto r^{1/2+\beta}$ for $f=1/2$: $|\tau_{\rm mig}|$ decreases as the planet migrates inwards. This explains why $\Ivlib / \Ivcross$ also increases with time. Similarly, in the alternative model of Section~\ref{sec:alternative_1D_model}, so long as $t \ll \tau_{\rm vis}$, $\Ivlib / \Ivcross$ is given by Eq.~(\ref{eq:clem_limit2}) and smoothly increases with time as the planet migrates inwards. It will change behaviour only if $t \gg \tau_{\rm vis}$. At the end of the simulations, $\tau_{\rm vis} / t$ does not go below $\approx 0.5$. The final value for $\Ivlib / \Ivcross$ is larger in the $h_0=0.07$ simulations than in the $h_0 = 0.05$ runs because the planet has migrated farther inwards.

    \section{Alternative figure comparing results of 2D and 1D simulations} \label{app:direct_compar}
\begin{figure*}[h!]
        \centering
        \includegraphics[trim={1cm 1.5cm 2cm 0},clip,width=\hsize]{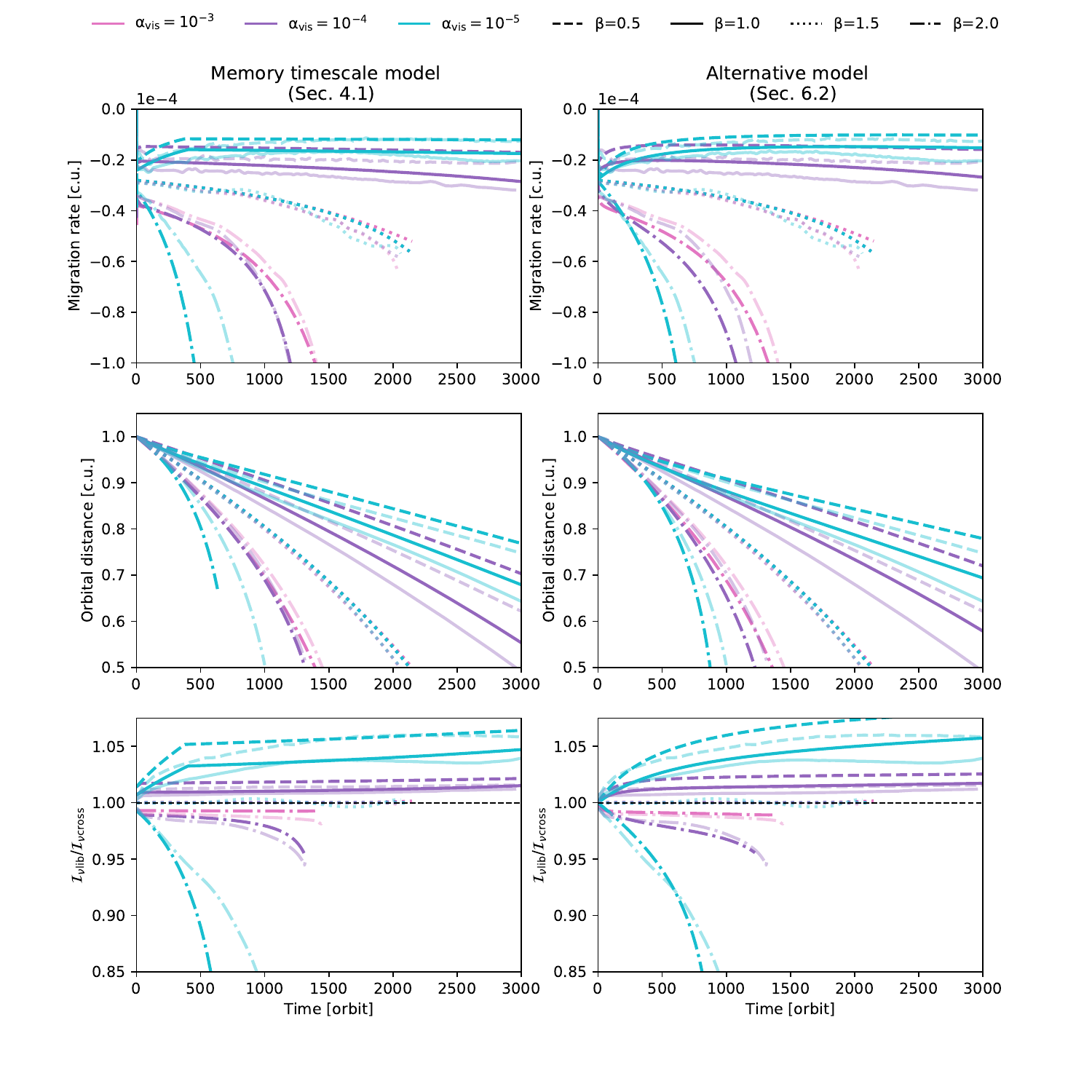}
        \caption{Overlay of 2D and 1D simulations results from Fig. \ref{fig:simulations_TQ8_migrate_orbital_evo} (memory timescale model of Section~\ref{sec:setup_1D}, left column) and Fig. \ref{fig:simulations_TQ8_migrate_evo_Clement} (alternative model of Section~\ref{sec:alternative_1D_model}, right column). Rows are the same as in Fig. \ref{fig:simulations_TQ8_migrate_orbital_evo}. Opaque lines display 1D simulations results, whereas transparent lines show the results of 2D hydrodynamical simulations. Cases with $\alpha_\mathrm{vis}=10^{-3}$ are only shown for $\beta \geq 1.5$ to make plots less crowded.}
        \label{fig:direct_comparison}
    \end{figure*}

    Throughout the manuscript, results of 2D and 1D simulations are compared for different 1D models of the dynamical corotation torque. A side-by-side comparison between 2D and 1D results is displayed in Figs. \ref{fig:simulations_TQ8_migrate_orbital_evo}, \ref{fig:non-isothermal_calculations}, \ref{fig:simulations_TQ8_migrate_evo_Clement} and \ref{fig:simulations_TQ8_migrate_evo_Paardekooper2014}. These figures also help assess the impact of including or not the dynamical corotation torque in 1D models. This appendix is used to include Fig.~\ref{fig:direct_comparison}, which provides a more direct comparison between the results of 2D and 1D simulations by overlying them on same panels. The left column shows the results of the memory timescale model in Fig.~\ref{fig:simulations_TQ8_migrate_orbital_evo}, the right column those of the alternative model in Fig.~\ref{fig:simulations_TQ8_migrate_evo_Clement}.

\end{appendix}

\end{document}